\documentclass[prd,aps,10pt,superscriptaddress,twocolumn,amsmath,amssymb,nofootinbib,longbibliography,floatfix]{revtex4-1}

\usepackage{ragged2e}

\usepackage[english]{babel}

\usepackage{natbib}
\usepackage[colorlinks]{hyperref}
\hypersetup{linkcolor=blue,citecolor=blue,filecolor=blue,urlcolor=blue}

\usepackage{amssymb,amsmath,multirow}
\usepackage{bm}
\usepackage{dsfont}
\usepackage{orcidlink}

\usepackage[utf8]{inputenc}
\usepackage{times}
\usepackage{setspace}
\usepackage{floatrow}
\usepackage{caption}
\usepackage[normalem]{ulem}

\usepackage{subfig,hyperref}

\usepackage{graphicx}
\usepackage{dcolumn}
\usepackage{bm}
\usepackage{wasysym}
\usepackage[
   starfontserif 
    ]{starfont}
\usepackage{amsmath}

\DeclareSymbolFont{starfontsym}{OT1}{sts}{m}{n}
\DeclareMathSymbol{\mathTerra}{\mathord}{starfontsym}{76}

\newcommand{\beq}{\begin{equation}}
\newcommand{\beqn}{\begin{align}}
\newcommand{\eeq}{\end{equation}}
\newcommand{\eeqn}{\end{align}}
\begin{document}

\title{Ultra-High-Energy Neutrinos from Primordial Black Holes}

\author{Alexandra P.~Klipfel \orcidlink{0000-0002-1907-7468}}

 \email{aklipfel@mit.edu}
\affiliation{%
Department of Physics, Massachusetts Institute of Technology, Cambridge, MA 02139, USA }

\author{David I.~Kaiser \orcidlink{0000-0002-5054-6744}}
 \email{dikaiser@mit.edu}
\affiliation{%
Department of Physics, Massachusetts Institute of Technology, Cambridge, MA 02139, USA
}%

\date{\today}

\begin{abstract}
The KM3NeT Collaboration recently announced the detection of a neutrino with energy 220 PeV. One possible source of such ultra-high-energy particles is the rapid emission of energetic Hawking radiation from a primordial black hole (PBH) near the end of its evaporation lifetime. The mass distribution for PBHs features a power-law tail for small masses; a small subset of PBHs would be undergoing late-stage evaporation today. We find that recent high-energy neutrino events detected by the IceCube and KM3NeT Collaborations, with energies ${\cal O} (1 - 10^2) \, {\rm PeV}$, are consistent with event-rate expectations if a significant fraction of the dark matter consists of PBHs.
\end{abstract}

\maketitle

{\it Introduction.} We investigate extremely energetic Hawking radiation from exploding primordial black holes (PBHs) as a possible source of ultra-high-energy cosmic rays, such as the KM3-230213A neutrino event recently reported by the KM3NeT Collaboration \cite{aiello_observation_2025}. The median neutrino energy required to produce the $120_{-60}^{+110}$ PeV muon observed by the KM3NeT ARCA detector on 13 February 2023 is 220 PeV, with a $68\%$ confidence interval of 110-790 PeV. The KM3-230213A event is the highest-energy neutrino event ever reported, although the IceCube Collaboration has reported detection of several PeV cosmic neutrinos since 2011 with deposited energies in the detector within the range $1.01 \pm 0.16 \, {\rm PeV} \leq E_{\rm dep} \leq 6.05\pm 0.72 \,{\rm PeV}$ \cite{icecube_collaboration_first_2013,icecube_collaboration_observation_2014,aartsen_observation_2016,icecube_collaboration_detection_2021}. Whereas an IceCube neutrino event with $E_{\rm dep} \sim {\cal O} (10^2 \, {\rm TeV})$ has been identified with emission from a blazar \cite{IceCube:2018cha,IceCube:2018dnn}, these rare, higher-energy events with $E_{\rm dep} \sim {\cal O} (1 - 10^2) \, {\rm PeV}$ have not been conclusively linked to any astrophysical point sources \cite{km3net_collaboration_characterising_2025}.

As analyzed in Refs.~\cite{li_clash_2025,Neronov:2025jfj}, to minimize tension with IceCube data, the ultra-high-energy KM3NeT neutrino most likely arose from a transient astrophysical 
point source, rather than from the diffuse isotropic neutrino flux. In this article we consider whether Hawking radiation from PBHs that are undergoing late-stage evaporation could account for rare, high-energy neutrino events. In particular, we investigate whether the small set of neutrinos detected by IceCube and KM3NeT with $E_{\rm dep} \sim {\cal O} (1 - 10^2 \, {\rm PeV})$, listed in Table~\ref{table:NuEvents}, 
could all have arisen from a single PBH distribution. (See Supplemental Materials \cite{SMnote}.)

PBHs were first considered as a theoretical possibility more than fifty years ago \cite{zeldovich_hypothesis_1966,hawking_gravitationally_1971,carr_black_1974}. Since PBHs would move at nonrelativistic speeds and interact gravitationally, they have emerged in recent years as an intriguing dark matter (DM) candidate. Recent work has clarified that PBHs could constitute most or all of the dark matter if their typical mass $\bar{M}$, at the peak of the mass distribution, falls within the range $10^{17} \, {\rm g} \leq \bar{M} \leq 10^{23} \, {\rm g}$ \cite{Khlopov:2008qy,carr_primordial_2020,carr_constraints_2021,carr_primordial_2022,green_primordial_2021,Kim:2020ngi,Berteaud:2022tws,carr_observational_2024,escriva_primordial_2024,gorton_how_2024,DelaTorreLuque:2024qms}. This window is referred to as the ``asteroid-mass'' range, which is exponentially smaller than a solar mass ($M_\odot = 2 \times 10^{33} \, {\rm g}$).

The lower bound on the asteroid-mass range, $\bar{M} \geq 10^{17} \, {\rm g}$, comes from consideration of PBH evaporation and cosmic ray measurements. Like any black hole, a PBH is expected to emit a thermal spectrum of Hawking radiation with an effective temperature $T_H$ inversely proportional to its mass \cite{hawking_black_1974,hawking_particle_1975}. The primary emission spectrum should consist of all elementary particles with masses $m_i \lesssim T_H$. A PBH would therefore lose mass over time via Hawking evaporation, yielding a finite lifetime $\tau \propto M_i^3$, where $M_i$ is the initial PBH mass at the time of its formation \cite{page_particle_1976,page_particle_1976-1,page_particle_1977,macgibbon_quark-_1990,macgibbon_quark-_1991}. Black hole evaporation is a runaway process: as a black hole's mass falls and its temperature rises, it should emit particles with correspondingly higher energies, ending in a final burst (or ``explosion'') in which particles would emerge with energies presumably approaching the Planck scale, $M_{\rm pl} \equiv 1 / \sqrt{ 8 \pi G} = 2.43 \times 10^{18} \, {\rm GeV}$.

We first describe how the small number of neutrino events detected by IceCube and KM3NeT with deposited energy $E_{\rm dep} \geq 1 \, {\rm PeV}$ could arise from a population of PBHs exploding in the Milky Way DM halo.
We find that the reported IceCube neutrino flux at 1 PeV, and therefore the five high-energy IceCube neutrino events with $E_{\rm dep} \sim {\cal O} (1) \, {\rm PeV}$, are consistent with an averaged, isotropic PBH explosion rate throughout the Milky Way Galaxy of $n_0 = 1.41^{+0.80}_{-0.71} \times 10^3 \, {\rm pc}^{-3} \, {\rm yr}^{-1}$ ($68\%$ confidence interval), which is compatible with the current most stringent upper bounds on $n_0$ from $\gamma$ ray and other high-energy cosmic ray searches \cite{glicenstein_limits_2013,abdo_milagro_2015,archambault_search_2017,the_fermi-lat_collaboration_search_2018,albert_constraining_2020}. Next we find that given this inferred value of $n_0$ from the IceCube events, there is a nontrivial probability to find at least one rare PBH explosion close enough to Earth to be consistent with the KM3-230213A neutrino event. We treat this ``local'' PBH explosion, which meets the transient point source criteria proposed by Ref.~\cite{li_clash_2025}, as a Poisson-distributed event drawn from the same underlying PBH distribution that sources the isotropic IceCube high-energy neutrino flux.

We then compute the expected explosion rate if PBHs constitute some fraction of all dark matter, based on a well-motivated PBH mass distribution. We identify regions of parameter space for values of $\bar{M} \sim 10^{17} \, {\rm g}$, near the lower end of the open asteroid-mass range, within which the expected PBH explosion rates are compatible (within $\sim 2 \sigma$) with those inferred by our analysis of the IceCube and KM3NeT neutrino fluxes. This congruence suggests a possible explanation for as-yet unexplained rare high-energy neutrino events.

{\it High-Energy Neutrinos from Hawking Radiation.} 
PBHs within the asteroid-mass range that arise from collapse of inflationary-era overdensities typically form with vanishing spin and charge 
\cite{chiba_spin_2017, de_luca_initial_2019, de_luca_evolution_2020, chongchitnan_extreme-value_2021,alonso-monsalve_primordial_2024}, so we consider Hawking emission from Schwarzschild black holes \cite{ChargeSpinNote}. Whether and how the semi-classical Hawking-radiation formalism might need to be modified at late stages of black hole evaporation remains an open question \cite{Dvali:2018xpy,Dvali:2020wft,zantedeschi_ultralight_2025,boccia_strike_2025}. As a conservative analysis, we work with the standard formalism for Hawking radiation \cite{hawking_black_1974,hawking_particle_1975,page_particle_1976,page_particle_1976-1,page_particle_1977,macgibbon_quark-_1990,macgibbon_quark-_1991}, updated to include the present-day set of Standard Model (SM) degrees of freedom.

The \textit{primary emission} rate per particle degree of freedom from a Schwarzschild black hole of mass $M$ is \cite{page_particle_1976, macgibbon_quark-_1990}
\begin{equation}
    \label{eqn:PrimarySpectraSchwarzschild}
    \frac{d^2N_s^{(1)}}{dtdQ} = \frac{\Gamma_s}{2 \pi \hbar} \left[\exp{\left( \frac{Q}{\hbar c^3/8\pi G M} \right)} - (-1)^{2s}  \right]^{-1}
\end{equation}
for emitted particles with spin $s$ and energy $Q$. In Table~\ref{table:DOF}, 
we list relevant parameters for all SM particles. The dynamics of scattering the field off the black hole potential are captured in the greybody factor $\Gamma_s$~\cite{page_particle_1976, macgibbon_quark-_1990, teukolsky_perturbations_1974, teukolsky_perturbations_1973}. The bracketed term on the righthand side of Eq. (\ref{eqn:PrimarySpectraSchwarzschild}) has the form of a blackbody spectrum for an emitter at temperature $T_H = \hbar c^3 / [8 \pi G M]$.

The PBH lifetime is governed by primary Hawking emission. The mass of a PBH obeys \cite{macgibbon_quark-_1991} 
\begin{equation}
\begin{split}
    \label{eqn:MassEvolution}
    \frac{dM}{dt} &= -\sum_j g_j\int_{m_{\rm eff}, j}^{\infty}dQ \, \frac{Q}{c^2}\frac{d^2N^{(1)}_{s_j}}{dtdQ}\\
    &=-5.34\times10^{25} \, \frac{f(M)}{M^2}\text{ g s}^{-1} ,
    \end{split}
\end{equation}
where $g_j$ counts the degrees of freedom per particle species, and $f (M)$ is known as the Page factor. We follow the method of Ref.~\cite{macgibbon_quark-_1991} to compute $f(M)$, but use the updated particle masses listed in Table~\ref{table:DOF} 
and include terms for the Higgs, $W$, and $Z$ bosons, which were not included in the original work. The Page factor can be computed via: 
\begin{equation}
    \label{eqn:PageFactor}
    f(M) \simeq \mathcal{P}_{\gamma}+\mathcal{P}_{\nu}+\sum_j \mathcal{P}_j\exp{\left(-\frac{M}{M_j} \right)},
\end{equation}
where the Page coefficients $\mathcal{P}_j$ are listed in column 8 of Table~\ref{table:DOF}, 
the characteristic PBH masses $M_j$ are listed in column 6, and the sum runs over the remaining 15 Standard Model particles. We consider only SM particles here. Including hypothetical Beyond-SM particles could either raise or lower the integrated number of high energy neutrinos emitted in model-dependent ways \cite{Perez-Gonzalez:2025try,BSMnote}. 

Because the exploding PBHs we consider are so light and hot, we must consider the net \textit{secondary emission} rates for SM particles, which are modified by particle decays, jet fragmentation, and hadronization \cite{macgibbon_quark-_1990, macgibbon_quark-_1991}. We denote secondary spectra as $d^2N^{(2)}/dtdQ$. The secondary spectra are computed numerically with \texttt{BlackHawk v2.2} \cite{arbey_physics_2021, arbey_blackhawk_2019} and \texttt{HDMSpectra} \cite{Bauer:2020jay}. 

For PBHs that form via the well-studied process of ``critical collapse'' \cite{gundlach_critical_2007}, the peak of the PBH mass distribution $\bar{M}$ scales with the mass enclosed within a Hubble sphere (known as the ``horizon mass'' $M_H$) at formation time $t_i$: $\bar{M} (t_i) = \eta \, M_H (t_i)$, with $\eta \simeq 0.2$ \cite{carr_black_1974,gundlach_critical_2007, Escriva:2021aeh,escriva_primordial_2024}. Assuming standard $\Lambda$CDM cosmological evolution since the end of cosmic inflation, we may then use the evolution of $M_H (t)$ to infer that a population of PBHs that formed within the asteroid-mass range, with $10^{17} \, {\rm g} \leq \bar{M} \leq 10^{23} \, {\rm g}$, formed at times $10^{-21} \, {\rm s} \leq t_i \leq 10^{-16} \, {\rm s}$ after the end of inflation \cite{alonso-monsalve_primordial_2024}. PBHs in our Galaxy today would therefore have been evolving over the entire age of the observable Universe: $t_0 = (13.787\pm0.020)\times10^9$ yr \cite{planck_collaboration_planck_2020}. Additionally, for PBHs with $M \ll M_\odot$, accretion remains exponentially suppressed, with $\Delta M / M_i < {\cal O} (10^{-2})$ between $t_i$ and $t_0$ \cite{rice_cosmological_2017, de_luca_constraints_2020, de_luca_accretion_2024}, so we neglect accretion when considering PBH evolution. (Here $M_i \equiv M (t_i)$.)

Numerically integrating Eq.~(\ref{eqn:MassEvolution}) allows us to solve for the PBH mass at formation time that corresponds to a lifetime equal to the current age of the Universe, $t_0$. This yields the cutoff mass $M_* \simeq 5\times10^{14}\text{ g}.$ In other words, a PBH that formed with $M_i = M_*$ would be undergoing its final evaporation process and exploding today. Treating the Page factor as varying slowly with $M$, the lifetime of a PBH that forms with mass $M_i$ may be approximated as 
\cite{macgibbon_quark-_1991}
  $  \tau(M_i) \simeq 1.98\times10^{-34}M_i^3f(M_i)^{-1} \text{ yr}$,
with $M_i$ measured in grams.

We compute the total number of emitted neutrinos with $Q\geq Q_{\rm min}$ produced during a PBH explosion. We consider two cases: $Q_{\rm min} = 1 \, {\rm PeV}$, the typical energy scale of the IceCube high-energy neutrino events, and $Q_{\rm min} = 60 \, {\rm PeV}$, which corresponds to the lower bound on the 68\% CI for the KM3-230213A event deposition. See Table~\ref{table:PBHInfo} for the relevant PBH parameters for each case.  

\begin{table}[h!]
\begin{ruledtabular}
\begin{tabular}{lccc}
$Q_{\rm peak}$ [PeV] & $M_0$ [g] & $T_0$ [PeV] & $\tau$ [s] \\
 \hline
1  & $4.79 \times 10^7$ & 0.221 & $4.57 \times 10^{-5}$ \\
60  & $7.98 \times 10^5$ & 13.2 & $2.21 \times 10^{-10}$ \\
\end{tabular}
\end{ruledtabular}
\caption{\justifying Parameters for PBHs producing neutrinos with energies above 1 PeV and 60 PeV, respectively. Here $\tau$ is the PBH's remaining lifetime when $M=M_0$. Note that we use $Q_{\rm peak} = 4.53 \, T_H$, appropriate for fermions \cite{macgibbon_quark-_1991}.}
\label{table:PBHInfo}
\end{table}

To estimate the number of emitted neutrinos with energies $E_\nu \geq 1 \, {\rm PeV}$, we make a well-motivated assumption that all particles emitted during the explosion of a PBH of mass $M_0 \lesssim 10^7$g have energy $Q\approx Q_{\text{peak}}(M_0)$. This is justified because the PBH lifetime---the time required for a PBH of mass $M$ to convert all its mass to Hawking radiation---scales as $\tau(M) \propto M^3$. The fraction of time a PBH spends with mass below some mass $M$ is therefore highly suppressed compared to the lifetime of reference mass $M_0 > M$. For example, by the time a PBH evaporates to have mass $M_0 = 4.79 \times 10^7 \, {\rm g}$, it will spend $90\%$ of its {\it remaining} lifetime with mass in the range $2.22 \times 10^7 \, {\rm g} \leq M \leq 4.79 \times 10^7 \, {\rm g}$, corresponding to Hawking emission with $1.00 \,  \text{ PeV} \leq Q_{\text{peak}} \leq 2.16 \, {\rm PeV}.$

\begin{figure}[h!]
    \centering
\includegraphics[width=0.99\linewidth]{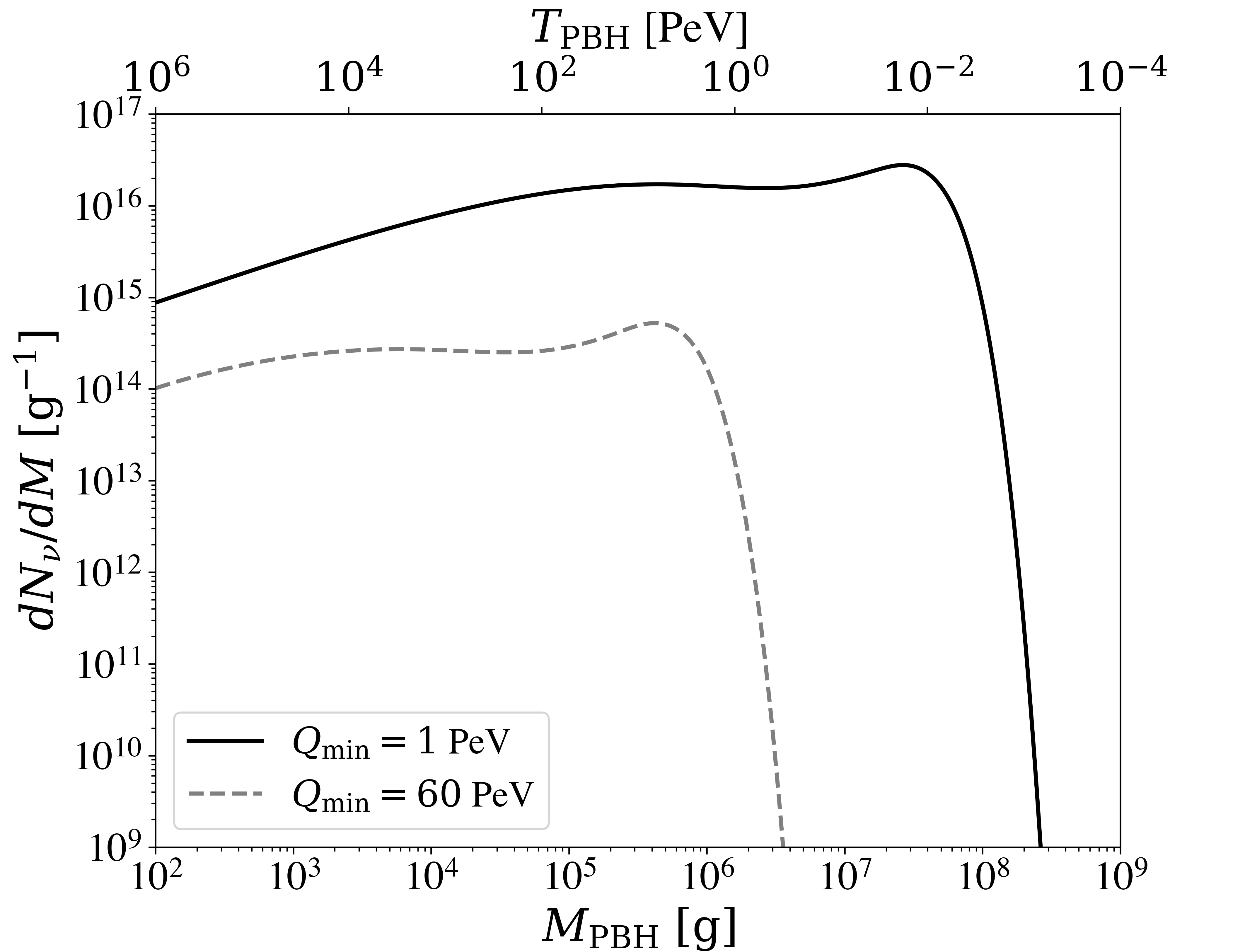}
    \caption{\justifying Differential emission rate for neutrinos with energies $Q\geq Q_{\rm min}$ during a PBH explosion. The mass $M_{\rm PBH}$ refers to the PBH mass at the time of particle emission. Note that neutrino emission rates peak for $M_{\rm PBH}\simeq M_0$ for $M_0$ values listed in Table~\ref{table:PBHInfo}.}
    \label{fig:PrimarySpectrum}
\end{figure}

Under this simplifying assumption, we may estimate the total number of emitted neutrinos with $Q\geq Q_{\rm min}$ during the PBH's final explosion. 
We first integrate over the secondary spectrum:
\beq
\frac{ d N_\nu (M, Q_{\rm min}) }{ dt} = \int_{Q_{\rm min}}^\infty dQ \, \frac{ d^2 N_{\nu}^{(2)} (M, Q)}{dt dQ} .
\label{dNdt}
\eeq
Then we use the PBH mass-loss rate from Eq.~(\ref{eqn:MassEvolution}) to evaluate
\beq
N_{\nu} (Q_{\rm min}) = \int_{10^0 \, {\rm g}}^{10^{10} \, {\rm g}} dM \, \frac{ d N_\nu (M, Q_{\rm min})}{dt} \left( \frac{ dM}{dt} \right)^{-1} .
\label{Nnutotal}
\eeq
Integrating up to $M \leq 10^{10} \, {\rm g}$ suffices, given how sharply the integrand falls off for $M > M_0$  (see Fig.~\ref{fig:PrimarySpectrum}). We compute this integral numerically, interpolating $d N_\nu (M, Q_{\rm min}) / dt$ over a list of values in our desired mass range generated numerically from secondary emission spectra. The results for $N_\nu$ are shown in Table~\ref{table:NuEvents}, 
indicating $N_\nu \sim 10^{24}$ for $Q_{\rm min} = 1 \, {\rm PeV}$ while $N_\nu \sim 10^{20}$ for $Q_{\rm min} = 60 \, {\rm PeV}$. In each case, we calculate the total number summed over all three neutrino flavors and their antiparticles because the IceCube and KM3NeT detectors are sensitive to all three neutrino flavors \cite{BallisticNote}.

{\it Inferred PBH Explosion Rate from High-Energy IceCube Neutrino Flux.}
Ref.~\cite{naab_measurement_2023} reports that the detected IceCube flux for each neutrino flavor for energies above $13.7 \, {\rm TeV}$ is best fit with a broken power law, parameterized as $\Phi_\nu^{\rm IC} (E_\nu) = \Phi_b (E_\nu / E_{\rm break} )^{- \gamma_i}$, where $\log_{10} (E_{\rm break} / {\rm GeV}) = 4.39^{+0.09}_{-0.08}$ and $\gamma_1 = 1.31^{+0.50}_{-1.21}$ ($E < E_{\rm break}$), $\gamma_2 = 2.74^{+0.06}_{-0.07}$ ($E \geq E_{\rm break}$) ($68\%$ confidence intervals). Given the reported flux at $E_\nu = 100 \, {\rm TeV}$ \cite{naab_measurement_2023}, this corresponds to $\Phi_b = 83.0_{-43.0}^{+48.2}$ per neutrino flavor, in units of $10^{-18} \, {\rm GeV}^{-1} \, {\rm cm}^{-2} \, {\rm sr}^{-1} \, {\rm s}^{-1}$ \cite{TestNote}. We may calculate the corresponding volumetric PBH explosion rate $n_0$ that would yield such a flux.

As discussed in the Supplemental Material, 
neutrino events with $E_\nu \geq 1 \, {\rm PeV}$ that originate from PBH explosions outside the Milky Way Galaxy would contribute a subdominant fraction to the expected event rate, so we focus on PBH explosions within our Galaxy. We adopt a modified Navarro-Frenk-White (NFW \cite{navarro_structure_1996}) parameterization for the spatial distribution $\rho_{\rm DM} (r,z)$ of the Milky Way DM halo \cite{binney_modelling_2017,posti_mass_2019}:
\begin{equation} 
    \label{eqn:DMProfile}
    \rho_{\text{DM}}(r, z) = \frac{\rho_0}{L(1+L)^2} \exp\left[-\left(\frac{L r_0}{r_{\rm vir}}\right)^2 \right],
\end{equation}
where $r$ and $z$ are cylindrical coordinates in the Galactocentric frame, $L \equiv [ (r / r_0)^2 + (z / [q r_0])^2 ]^{1/2}$, and best-fit parameters are $\rho_0 = 0.0196 \, M_\odot \, {\rm pc}^{-3}$, $r_0 = 15.5 \, {\rm kpc}$, $r_{\rm vir} = 287 \, {\rm kpc}$, and $q = 1.22$. The PBH explosion rate $n_0$ would then scale with the DM density profile as 
\begin{equation}
    n_0(\vec{r}|n_{\rm NFW})=\frac{n_{\rm NFW}}{\rho_0}\rho_{\text{DM}}(r, z),
 \label{n0rhoDM}
\end{equation}
where $n_{\rm NFW}$ is a normalization constant with units of ${\rm pc}^{-3} \, {\rm yr}^{-1}$. Note that we have made no assumption about the fraction $f_{\rm PBH}$ of DM contributed by PBHs; we assume only that the spatial distribution of any exploding PBHs that might exist tracks the overall DM density profile. Furthermore, we do not make any assumptions about the underlying PBH mass function. We neglect any potential effects due to PBH interactions, clustering, or binaries \cite{carr_observational_2024}. 

The isotropic flux of neutrinos with 
$E_\nu \geq Q_{\rm min}$ at Earth is given by a volume integral over the DM halo,
\begin{equation}
    \label{eqn:ExplosionFlux}
    \Phi_{\nu}(n_{\rm NFW}, Q_{\rm min})=\frac{1}{4\pi Q_{\rm min}} \int_V  \frac{dV \, N_{\nu} (Q_{\rm min})}{4\pi(\vec{r}-\vec{R}_{\astrosun})^2}n_0(\vec{r}|n_{\rm NFW}),
\end{equation}
where $\Phi_{\nu}$ has dimensions ${\rm GeV}^{-1}{\rm cm}^{-2}\, {\rm s}^{-1}\, {\rm sr}^{-1}\, $, $R_{\astrosun} = 8.3$ kpc is the location of the Solar System \cite{posti_mass_2019}, and we take $V$ to be a sphere of radius $r_{\rm vir}$ centered at the center of the Milky Way. Upon solving for the normalization constant $n_{\rm NFW}$ such that $Q_{\rm min}=1$ PeV and $\Phi_\nu (n_{\rm NFW}, Q_{\rm min}) = 3\Phi_\nu^{\rm IC} (Q_{\rm min})$, we then find the PBH explosion rate in the neighborhood of our Solar System to be $n_0 =  n_{\rm NFW}  \, [\rho_{\rm DM} (R_\odot, 0) / \rho_0 ] = 1.41^{+0.80}_{-0.71} \times 10^3 \, {\rm pc}^{-3} \, {\rm yr}^{-1}$ ($68\% \, {\rm CI})$. (The wide error bars in $n_0$ arise largely from the uncertainty in $E_{\rm break}$ in the parameterized broken power law for the IceCube flux.) This inferred value of $n_0$ is compatible with the current most stringent upper bound $n_0 < 3.40^{+0.40}_{-0.10} \times 10^3 \, {\rm pc}^{-1} \, {\rm yr}^{-1}$ ($99\%$ CI) reported by the HAWC Collaboration based on $\gamma$-ray searches \cite{albert_constraining_2020}. Repeating the above analysis using the reported KM3NeT flux for 60 PeV neutrinos \cite{aiello_observation_2025}, we determine that the volumetric explosion rate necessary to source the KM3NeT isotropic flux is $n_0=1.48_{-0.47}^{+1.29}\times10^6 \, {\rm pc}^{-3}\, {\rm yr}^{-1}$, which is approximately $10^3$ times larger than the IceCube result. These inferred explosion rates are therefore in tension at $\gtrsim 3 \sigma$, as are the reported isotropic fluxes, as noted by Ref.~\cite{li_clash_2025}. 

To resolve this tension, we use the inferred value of $n_0$ from the isotropic IceCube neutrino flux at 1 PeV to estimate the probability for at least one PBH explosion to occur within some distance $b$ of the Earth during an observing window $T$. We assume that such a rare event would be governed by Poisson statistics: ${\rm Prob} (\text{1 event in $T$}) = (r T) \exp [ - rT]$, where $r$ is the rate. During an observing window $T$, the volume to which Earthbound detectors would be sensitive as the Earth moves through the DM halo is $V (b, T) = (4 \pi / 3) b^3 + \pi b^2 v_{\rm avg} T$.  We estimate $v_{\rm avg}$, the average relative velocity between the Earth and the local DM, by averaging over a Maxwellian velocity distribution with $v_{\rm rms} = 270 \, {\rm km / s}$ up to the escape velocity $v_{\rm esc} = 544 \, {\rm km / s}$ \cite{cerdeno_particle_2010,choi_impact_2014}, which yields $v_{\rm avg} = 246 \, {\rm km / s}$. In order for at least one ultra-high-energy neutrino with $E_\nu \geq 60 \, {\rm PeV}$ to strike some area $A$ on Earth, we require $b_{\rm max} = [ N_\nu A / (4 \pi) ]^{1/2}$. Using $N_\nu (60 \,{\rm PeV}) = 4.02 \times 10^{20}$ from Table~\ref{table:NuEvents}, 
$A = (50 \, {\rm km})^2$ and $r=n_0 V(b_{\rm max}, T=14 {\rm yr})$, we find ${\rm Prob} (\text{1 event since 2011}) = 0.076$. This yields a nontrivial likelihood to have one detectable PBH explosion within $b_{\rm max} \simeq 1890 \, {\rm AU}$ since 2011.

{\it PBH Distribution and Expected Explosion Rates.} We follow Refs.~\cite{mosbech_effects_2022,Cang:2021owu} and define the PBH number distribution at the time of formation $t_i$,
\beq
\phi (M_i) = \frac{1}{n_{\rm PBH} } \frac{ dn}{dM_i} ,
\label{phidef}
\eeq
where 
$n_{\rm PBH} = f_{\rm PBH} \, \rho_{\rm DM} / \int_0^\infty dM_i \, M_i \, \phi (M_i)$ is the total number density of PBHs that form at $t_i$. The initial number distribution is normalized: $\int_0^\infty dM_i \, \phi (M_i) = 1$. 
As individual PBHs undergo Hawking emission, their masses will change over time. Assuming that the Page factor $f (M)$ varies slowly over time \cite{PageNote}, we may solve Eq.~(\ref{eqn:MassEvolution}) for the mass $M(t \vert M_i)$ at some time $t > t_i$ \cite{mosbech_effects_2022}:
\beq
M (t \vert M_i) = \left( M_i^3 \frac{ f (M)}{f (M_i)} - 3 \tilde{\alpha} t \, f (M) \right)^{1/3} ,
\label{Mt}
\eeq
where $\tilde{\alpha} = 5.34 \times 10^{25} \, {\rm g}^3 \, {\rm s}^{-1}$. The interval $dM$ and the function $\phi (M, t)$ will also evolve over time. Again taking $f(M)$ to vary slowly, we have $
\phi (M, t) \simeq M^2 \phi (M_i ) \left( M^3 + 3 f(M) \tilde{\alpha} t  \right)^{-2/3}$ \cite{mosbech_effects_2022}.
The fraction of PBHs that remain at time $t > t_i$ is
$F (t) \equiv \int_0^\infty dM \, \phi (M, t)$,
and $1 - F (t)$ is the cumulative fraction of PBHs that have completely evaporated by time $t$. To compute the instantaneous PBH explosion rate per unit volume, we therefore write
\beq
\dot{N} (t) = n_{\rm PBH} \frac{d}{dt} \left( 1 - F (t) \right) .
\label{dotNdef}
\eeq
We use $\dot{N} (t)$ to compute the average volumetric PBH explosion rate $\cal{N}$ within a time interval $[ t_0 - \Delta T, t_0]$ within the neighborhood of our Solar System. For a given PBH number distribution $\phi (M \vert \{ \theta_j \} )$ that depends on some set $\{ \theta_j \}$ of $j$ tunable parameters, we evaluate
\beq
{\cal N} ( \{ \theta_j \} ) = \frac{1}{\Delta T} \int_{t_0 - \Delta T}^{t_0} dt \, \dot{N} (t \vert \{ \theta_j \} ) \simeq \dot{N} (t_0 \vert \{ \theta_j \} ) ,
\label{calNdef}
\eeq
where the last step follows since we are interested in durations $\Delta T \sim {\cal O} (10 \, {\rm yr})$ compared to $t_0 \sim {\cal O} (10^{10} \, {\rm yr})$.

The quantity ${\cal N}$ is the time-averaged PBH explosion rate per unit volume with units of ${\rm pc}^{-3} \, {\rm yr}^{-1}$, which depends on the underlying PBH number distribution $\phi (M, t)$, which in turn depends on $n_{\rm PBH}$ as defined below Eq.~(\ref{phidef}). Given that $\rho_{\rm DM} (r, z)$ in Eq.~(\ref{eqn:DMProfile}) is nearly constant over length-scales $r \leq {\cal O} (1 \, {\rm kpc})$ and we are considering distances of order $b_{\rm max} \sim {\cal O} (10^3 \, {\rm AU}) \sim {\cal O} (10^{-5} \, {\rm kpc})$, we may set $\rho_{\rm DM} = \rho_{\odot} = 0.0155 \, M_\odot \, {\rm pc}^{-3}$. Present evaporation constraints limit $f_{\rm PBH} < 1$ for mass functions that peak with $\bar{M} \lesssim 10^{17}$ g \cite{carr_primordial_2020,carr_constraints_2021,carr_primordial_2022,green_primordial_2021,Kim:2020ngi,Berteaud:2022tws,carr_observational_2024,escriva_primordial_2024,gorton_how_2024,DelaTorreLuque:2024qms}. From the parameterization in Ref.~\cite{DelaTorreLuque:2024qms} of the most recent 511 MeV $\gamma$-ray constraints reported by Ref.~\cite{Berteaud:2022tws}, the endpoints of the ``realistic'' $3 \sigma$ confidence-interval within which $f_{\rm PBH} = 1$ lie within the range $(1.5 -2.5) \times10^{17} \, {\rm g}$. We use the representative value of $2.0 \times10^{17} \, {\rm g}$ and set $f_{\rm PBH} \simeq (\bar{M}/ [2 \times 10^{17} \, {\rm g} ])^3$ for $\bar{M} \leq 2\times  10^{17} \, {\rm g}$.

Decades of detailed numerical-relativity simulations indicate that PBHs form via a process known as ``critical collapse.'' (For reviews, see Refs.~\cite{gundlach_critical_2007,Escriva:2021aeh}.) Critical collapse yields a nontrivial mass distribution $\psi (M_i)$ for the resulting PBH population that is sharply peaked at a value $\bar{M}$, with a power-law tail for masses $M_i < \bar{M}$ and an exponential fall-off for $M_i > \bar{M}$ \cite{Niemeyer:1997mt,Green:1999xm,Kuhnel:2015vtw}. As analyzed in Ref.~\cite{gow_accurate_2022}, the generalized critical collapse (GCC) mass distribution $\psi_{\rm GCC} (M_i)$  yields the closest fit to the PBH mass distributions arising from several known spectra of primordial curvature perturbations. The associated GCC number distribution $\phi_{\rm GCC} (M_i)$ may be written 
\beq
\phi_{\rm GCC} (M_i ) = \frac{ \beta}{\mu \, \Gamma \left(\frac{ \alpha }{\beta} \right)} \left( \frac{ M_i}{\mu} \right)^{\alpha -1}\exp \left[ - \left( \frac{ M_i}{\mu} \right)^\beta \right],
\label{PhiGCCdef}
\eeq
with $\beta > 0$ and $\alpha > 1$ \cite{gow_accurate_2022,PhiPsiNote}. The peak mass $\bar{M}$ is related to the parameter $\mu$ as $\bar{M} = \mu [ (\alpha -1 ) / \beta]^{1 / \beta}$.

\begin{figure*}
\centering
\includegraphics[width=0.81\linewidth]{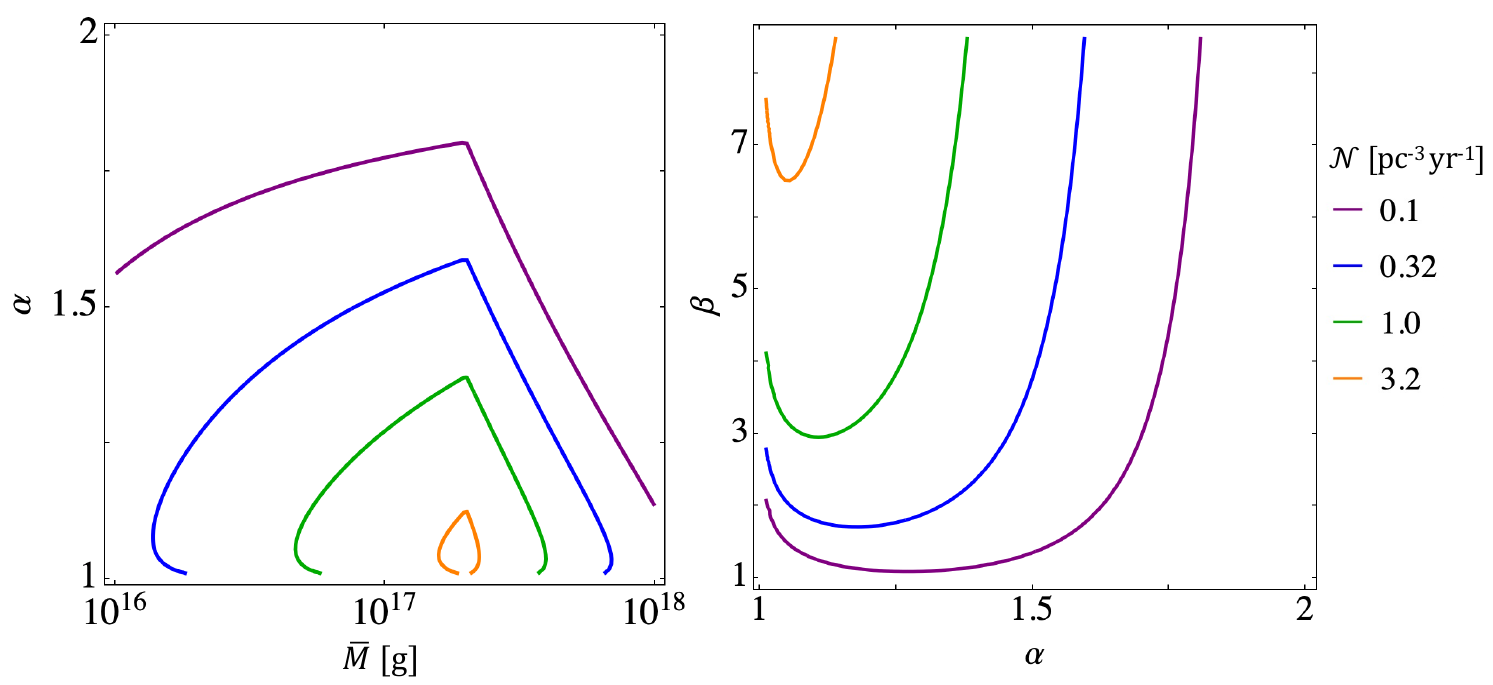}
\caption{\justifying  
Predicted volumetric PBH explosion rate ${\cal N}$ from the PBH number distribution $\phi_{\rm GCC} (M_i)$ of Eq.~(\ref{PhiGCCdef}) as a function of $\bar{M}$ and $\alpha$ with $\beta = 8$ ({\it left}) and as a function of $\alpha$ and $\beta$ with $\bar{M} = 2\times10^{17} \, {\rm g}$ ({\it right}). Note that a predicted rate of ${\cal N} \geq 1 \, {\rm pc}^{-3} \, {\rm yr}^{-1}$ (green curves) is within $1.98 \, \sigma$ of the mean value for $n_0$ inferred from the IceCube data, given the residual uncertainty on $E_{\rm break}$ in the parameterized high-energy IceCube neutrino flux. When calculating ${\cal N}$ we incorporate present constraints on $f_{\rm PBH}$ for $\bar{M} < 2 \times 10^{17} \, {\rm g}$.}
\label{fig:n0Mbar}
\end{figure*}

In Fig.~\ref{fig:n0Mbar} we show the expected present-day isotropic volumetric explosion rate ${\cal N} (\{ \theta_j \} )$ for the GCC distribution. To account for the neutrino events with $E_{\rm dep} \geq 1 \, {\rm PeV}$ in a way that is both consistent with the reported IceCube high-energy neutrino flux and with the anomalous KM3NeT transient event, we require ${\cal N} \geq n_0 = 1.41^{+0.80}_{-0.71} \times 10^3 \, {\rm pc}^{-3} \, {\rm yr}^{-1}$.
We find narrow regions of parameter space $\{\bar{M}, \alpha, \beta\}$ for which the expected explosion rate ${\cal N}$ is within $\sim 2 \sigma$ of the mean value of $n_0$ inferred from the IceCube isotropic flux. For those same regions of parameter space, we find that the spectrum of lower-energy neutrinos (${\rm keV} \leq E_\nu \leq 10^2 \, {\rm TeV}$) remains comparable to but below the so-called ``Grand Unified Neutrino Spectrum'' expected from other sources \cite{Vitagliano:2019yzm,Ivanchik:2024mqq}.

{\it Discussion.} Over the past fifteen years, a small number of high-energy neutrino events have been detected by the IceCube and KM3NeT Collaborations with energies $E_\nu \geq 1 \, {\rm PeV}$. To date, no astrophysical acceleration mechanism has been identified that could account for these specific events. The isotropic fluxes and inferred 
explosion rates from these few IceCube and KM3NeT events are incompatible with each other \cite{li_clash_2025}. However, we have shown that if this collection of high-energy neutrino events arose from the late-stage evaporation of primordial black holes within the dark matter halo of the Milky Way Galaxy, then the recent ultra-high-energy KM3NeT neutrino event with 
$E_\nu \sim {\cal O}(100) \, {\rm PeV}$ {\it would} be compatible with the reported high-energy neutrino flux from IceCube. The KM3NeT event could have originated from a rare, nearby transient PBH explosion drawn from the same underlying PBH distribution that sources the IceCube isotropic neutrino flux for $E_{\nu}\gtrsim 1 {\rm PeV}$. Moreover, the inferred PBH explosion rate $n_0$ consistent with these detections is compatible with the current most stringent upper bound on $n_0$ set by the HAWC Collaboration, based on $\gamma$-ray searches \cite{albert_constraining_2020}.

Not only does the hypothesis that these high-energy neutrino events each arose from PBH explosions reduce the tension between the reported IceCube and KM3NeT fluxes; it also enables us to estimate {\it ab initio} the expected number of such events. If PBHs formed via critical collapse very early in cosmic history, the resulting population would necessarily follow a nontrivial mass distribution, including a power-law tail for masses below the peak of the distribution \cite{Niemeyer:1997mt,Green:1999xm,Kuhnel:2015vtw,gow_accurate_2022,gorton_how_2024}. Given the wide error bars on the inferred isotropic PBH explosion rate from the IceCube data (driven by the uncertainty in the energy $E_{\rm break}$ in the broken power-law parameterization of the high-energy IceCube flux), we find regions of parameter space in which a population of PBHs with $f_{\rm PBH} = 1$ would include 
a small fraction of PBHs that are undergoing their final Hawking evaporation and exploding today at a rate consistent (within $\sim 2 \sigma$) with the rate required to account for all reported neutrino events to date with $E_\nu \geq 1 \, {\rm PeV}$. 

At least three different inputs could test or constrain the scenario presented here, each coming from quite distinct research communities. First, additional high-energy neutrino events from IceCube and/or KM3NeT will reduce uncertainties on reported fluxes and further test the compatibility between expected PBH explosion rates and observed neutrino fluxes. Second, upcoming high-energy $\gamma$-ray detectors, such as the Large High Altitude Air Shower Observatory (LHAASO) \cite{Yang:2024vij}, could be sensitive to local PBH explosion rates as low as $n_0 \simeq 1200 \, {\rm pc}^{-3} \, {\rm yr}^{-1}$, comparable to the inferred rate from present-day IceCube data. And third, other astrophysical probes of the PBH asteroid-mass range, including gravitational perturbations \cite{Qin:2023lgo,Tran:2023jci,Cuadrat-Grzybowski:2024uph,DeLorenci:2025wbn} and cosmic rays \cite{Klipfel:2025bvh}, could further constrain the parameter space within which $f_{\rm PBH} \simeq 1$, thereby modifying expected PBH explosion rates.

With more neutrino detection events for $E_\nu \geq 1 \, {\rm PeV}$, we may further clarify whether higher event rates originate from the direction of the galactic center, where the dark matter density is greatest. Predicting the magnitude of such a directional anisotropy remains the subject of further research.

{\it Acknowledgements.} We are grateful to Michael Baker, Shyam Balaji, Bryce Cyr, Peter Fisher, Anne Green, Joaquim Iguaz Juan, Evan McDonough, Benjamin Lehmann, Dario Lorenzoni, Chris Schmidt, Aidan Symons, Thomas Steingasser, Hayden Tedrake, Jesse Thaler, Andrea Thamm, and Rainer Weiss for helpful discussions. This material is based upon work supported by the National Science Foundation Graduate Research Fellowship under Grant No.~2141064. Portions of this work were conducted in MIT's Center for Theoretical Physics -- A Leinweber Institute and supported in part by the U.~S.~Department of Energy under Contract No.~DE-SC0012567.


\begin{thebibliography}{91}%
\makeatletter
\providecommand \@ifxundefined [1]{%
 \@ifx{#1\undefined}
}%
\providecommand \@ifnum [1]{%
 \ifnum #1\expandafter \@firstoftwo
 \else \expandafter \@secondoftwo
 \fi
}%
\providecommand \@ifx [1]{%
 \ifx #1\expandafter \@firstoftwo
 \else \expandafter \@secondoftwo
 \fi
}%
\providecommand \natexlab [1]{#1}%
\providecommand \enquote  [1]{``#1''}%
\providecommand \bibnamefont  [1]{#1}%
\providecommand \bibfnamefont [1]{#1}%
\providecommand \citenamefont [1]{#1}%
\providecommand \href@noop [0]{\@secondoftwo}%
\providecommand \href [0]{\begingroup \@sanitize@url \@href}%
\providecommand \@href[1]{\@@startlink{#1}\@@href}%
\providecommand \@@href[1]{\endgroup#1\@@endlink}%
\providecommand \@sanitize@url [0]{\catcode `\\12\catcode `\$12\catcode `\&12\catcode `\#12\catcode `\^12\catcode `\_12\catcode `\%12\relax}%
\providecommand \@@startlink[1]{}%
\providecommand \@@endlink[0]{}%
\providecommand \url  [0]{\begingroup\@sanitize@url \@url }%
\providecommand \@url [1]{\endgroup\@href {#1}{\urlprefix }}%
\providecommand \urlprefix  [0]{URL }%
\providecommand \Eprint [0]{\href }%
\providecommand \doibase [0]{http://dx.doi.org/}%
\providecommand \selectlanguage [0]{\@gobble}%
\providecommand \bibinfo  [0]{\@secondoftwo}%
\providecommand \bibfield  [0]{\@secondoftwo}%
\providecommand \translation [1]{[#1]}%
\providecommand \BibitemOpen [0]{}%
\providecommand \bibitemStop [0]{}%
\providecommand \bibitemNoStop [0]{.\EOS\space}%
\providecommand \EOS [0]{\spacefactor3000\relax}%
\providecommand \BibitemShut  [1]{\csname bibitem#1\endcsname}%
\let\auto@bib@innerbib\@empty
\bibitem [{\citenamefont {Aiello}\ \emph {et~al.}(2025)\citenamefont {Aiello} \emph {et~al.}}]{aiello_observation_2025}%
  \BibitemOpen
  \bibfield  {author} {\bibinfo {author} {\bibfnamefont {S.}~\bibnamefont {Aiello}} \emph {et~al.} (\bibinfo {collaboration} {KM3NeT}),\ }\bibfield  {title} {\enquote {\bibinfo {title} {{Observation of an ultra-high-energy cosmic neutrino with KM3NeT}},}\ }\href {\doibase 10.1038/s41586-024-08543-1} {\bibfield  {journal} {\bibinfo  {journal} {Nature}\ }\textbf {\bibinfo {volume} {638}},\ \bibinfo {pages} {376--382} (\bibinfo {year} {2025})}\BibitemShut {NoStop}%
\bibitem [{\citenamefont {Aartsen}\ \emph {et~al.}(2013)\citenamefont {Aartsen} \emph {et~al.}}]{icecube_collaboration_first_2013}%
  \BibitemOpen
  \bibfield  {author} {\bibinfo {author} {\bibfnamefont {M.~G.}\ \bibnamefont {Aartsen}} \emph {et~al.} (\bibinfo {collaboration} {IceCube}),\ }\bibfield  {title} {\enquote {\bibinfo {title} {{First observation of PeV-energy neutrinos with IceCube}},}\ }\href {\doibase 10.1103/PhysRevLett.111.021103} {\bibfield  {journal} {\bibinfo  {journal} {Phys. Rev. Lett.}\ }\textbf {\bibinfo {volume} {111}},\ \bibinfo {pages} {021103} (\bibinfo {year} {2013})},\ \Eprint {http://arxiv.org/abs/1304.5356} {arXiv:1304.5356 [astro-ph.HE]} \BibitemShut {NoStop}%
\bibitem [{\citenamefont {Aartsen}\ \emph {et~al.}(2014)\citenamefont {Aartsen} \emph {et~al.}}]{icecube_collaboration_observation_2014}%
  \BibitemOpen
  \bibfield  {author} {\bibinfo {author} {\bibfnamefont {M.~G.}\ \bibnamefont {Aartsen}} \emph {et~al.} (\bibinfo {collaboration} {IceCube}),\ }\bibfield  {title} {\enquote {\bibinfo {title} {{Observation of High-Energy Astrophysical Neutrinos in Three Years of IceCube Data}},}\ }\href {\doibase 10.1103/PhysRevLett.113.101101} {\bibfield  {journal} {\bibinfo  {journal} {Phys. Rev. Lett.}\ }\textbf {\bibinfo {volume} {113}},\ \bibinfo {pages} {101101} (\bibinfo {year} {2014})},\ \Eprint {http://arxiv.org/abs/1405.5303} {arXiv:1405.5303 [astro-ph.HE]} \BibitemShut {NoStop}%
\bibitem [{\citenamefont {Aartsen}\ \emph {et~al.}(2016)\citenamefont {Aartsen} \emph {et~al.}}]{aartsen_observation_2016}%
  \BibitemOpen
  \bibfield  {author} {\bibinfo {author} {\bibfnamefont {M.~G.}\ \bibnamefont {Aartsen}} \emph {et~al.} (\bibinfo {collaboration} {IceCube}),\ }\bibfield  {title} {\enquote {\bibinfo {title} {{Observation and Characterization of a Cosmic Muon Neutrino Flux from the Northern Hemisphere using six years of IceCube data}},}\ }\href {\doibase 10.3847/0004-637X/833/1/3} {\bibfield  {journal} {\bibinfo  {journal} {Astrophys. J.}\ }\textbf {\bibinfo {volume} {833}},\ \bibinfo {pages} {3} (\bibinfo {year} {2016})},\ \Eprint {http://arxiv.org/abs/1607.08006} {arXiv:1607.08006 [astro-ph.HE]} \BibitemShut {NoStop}%
\bibitem [{\citenamefont {Aartsen}\ \emph {et~al.}(2021)\citenamefont {Aartsen} \emph {et~al.}}]{icecube_collaboration_detection_2021}%
  \BibitemOpen
  \bibfield  {author} {\bibinfo {author} {\bibfnamefont {M.~G.}\ \bibnamefont {Aartsen}} \emph {et~al.} (\bibinfo {collaboration} {IceCube}),\ }\bibfield  {title} {\enquote {\bibinfo {title} {{Detection of a particle shower at the Glashow resonance with IceCube}},}\ }\href {\doibase 10.1038/s41586-021-03256-1} {\bibfield  {journal} {\bibinfo  {journal} {Nature}\ }\textbf {\bibinfo {volume} {591}},\ \bibinfo {pages} {220--224} (\bibinfo {year} {2021})},\ \bibinfo {note} {[Erratum: Nature 592, E11 (2021)]},\ \Eprint {http://arxiv.org/abs/2110.15051} {arXiv:2110.15051 [hep-ex]} \BibitemShut {NoStop}%
\bibitem [{\citenamefont {Aartsen}\ \emph {et~al.}(2018{\natexlab{a}})\citenamefont {Aartsen} \emph {et~al.}}]{IceCube:2018cha}%
  \BibitemOpen
  \bibfield  {author} {\bibinfo {author} {\bibfnamefont {M.~G.}\ \bibnamefont {Aartsen}} \emph {et~al.} (\bibinfo {collaboration} {IceCube}),\ }\bibfield  {title} {\enquote {\bibinfo {title} {{Neutrino emission from the direction of the blazar TXS 0506+056 prior to the IceCube-170922A alert}},}\ }\href {\doibase 10.1126/science.aat2890} {\bibfield  {journal} {\bibinfo  {journal} {Science}\ }\textbf {\bibinfo {volume} {361}},\ \bibinfo {pages} {147--151} (\bibinfo {year} {2018}{\natexlab{a}})},\ \Eprint {http://arxiv.org/abs/1807.08794} {arXiv:1807.08794 [astro-ph.HE]} \BibitemShut {NoStop}%
\bibitem [{\citenamefont {Aartsen}\ \emph {et~al.}(2018{\natexlab{b}})\citenamefont {Aartsen} \emph {et~al.}}]{IceCube:2018dnn}%
  \BibitemOpen
  \bibfield  {author} {\bibinfo {author} {\bibfnamefont {M.~G.}\ \bibnamefont {Aartsen}} \emph {et~al.} (\bibinfo {collaboration} {IceCube, Fermi-LAT, MAGIC, AGILE, ASAS-SN, HAWC, H.E.S.S., INTEGRAL, Kanata, Kiso, Kapteyn, Liverpool Telescope, Subaru, Swift NuSTAR, VERITAS, VLA/17B-403}),\ }\bibfield  {title} {\enquote {\bibinfo {title} {{Multimessenger observations of a flaring blazar coincident with high-energy neutrino IceCube-170922A}},}\ }\href {\doibase 10.1126/science.aat1378} {\bibfield  {journal} {\bibinfo  {journal} {Science}\ }\textbf {\bibinfo {volume} {361}},\ \bibinfo {pages} {eaat1378} (\bibinfo {year} {2018}{\natexlab{b}})},\ \Eprint {http://arxiv.org/abs/1807.08816} {arXiv:1807.08816 [astro-ph.HE]} \BibitemShut {NoStop}%
\bibitem [{\citenamefont {Adriani}\ \emph {et~al.}(2025)\citenamefont {Adriani} \emph {et~al.}}]{km3net_collaboration_characterising_2025}%
  \BibitemOpen
  \bibfield  {author} {\bibinfo {author} {\bibfnamefont {O.}~\bibnamefont {Adriani}} \emph {et~al.} (\bibinfo {collaboration} {KM3NeT, MessMapp Group, Fermi-LAT, Owens Valley Radio Observatory 40-m Telescope Group, SVOM}),\ }\bibfield  {title} {\enquote {\bibinfo {title} {{Characterising Candidate Blazar Counterparts of the Ultra-High-Energy Event KM3-230213A}},}\ }\href@noop {} {\  (\bibinfo {year} {2025})},\ \Eprint {http://arxiv.org/abs/2502.08484} {arXiv:2502.08484 [astro-ph.HE]} \BibitemShut {NoStop}%
\bibitem [{\citenamefont {Li}\ \emph {et~al.}(2025)\citenamefont {Li}, \citenamefont {Machado}, \citenamefont {Naredo-Tuero},\ and\ \citenamefont {Schwemberger}}]{li_clash_2025}%
  \BibitemOpen
  \bibfield  {author} {\bibinfo {author} {\bibfnamefont {Shirley~Weishi}\ \bibnamefont {Li}}, \bibinfo {author} {\bibfnamefont {Pedro}\ \bibnamefont {Machado}}, \bibinfo {author} {\bibfnamefont {Daniel}\ \bibnamefont {Naredo-Tuero}}, \ and\ \bibinfo {author} {\bibfnamefont {Thomas}\ \bibnamefont {Schwemberger}},\ }\href@noop {} {\enquote {\bibinfo {title} {{Clash of the Titans: ultra-high energy KM3NeT event versus IceCube data}},}\ } (\bibinfo {year} {2025}),\ \Eprint {http://arxiv.org/abs/2502.04508} {arXiv:2502.04508 [astro-ph.HE]} \BibitemShut {NoStop}%
\bibitem [{\citenamefont {Neronov}\ \emph {et~al.}(2025)\citenamefont {Neronov}, \citenamefont {Oikonomou},\ and\ \citenamefont {Semikoz}}]{Neronov:2025jfj}%
  \BibitemOpen
  \bibfield  {author} {\bibinfo {author} {\bibfnamefont {Andrii}\ \bibnamefont {Neronov}}, \bibinfo {author} {\bibfnamefont {Foteini}\ \bibnamefont {Oikonomou}}, \ and\ \bibinfo {author} {\bibfnamefont {Dmitri}\ \bibnamefont {Semikoz}},\ }\bibfield  {title} {\enquote {\bibinfo {title} {{KM3-230213A: An Ultra-High Energy Neutrino from a Year-Long Astrophysical Transient}},}\ }\href@noop {} {\  (\bibinfo {year} {2025})},\ \Eprint {http://arxiv.org/abs/2502.12986} {arXiv:2502.12986 [astro-ph.HE]} \BibitemShut {NoStop}%
\bibitem [{SMn()}]{SMnote}%
  \BibitemOpen
  \href@noop {} {}\bibinfo {note} {See the Supplemental Materials (SM), which includes discussion of the expected flux from extragalactic PBH explosions as well as tables of cosmogenic high-energy neutrino events and Page coefficients for each Standard Model particle.}\BibitemShut {Stop}%
\bibitem [{\citenamefont {Zel'dovich}\ and\ \citenamefont {Novikov}(1966)}]{zeldovich_hypothesis_1966}%
  \BibitemOpen
  \bibfield  {author} {\bibinfo {author} {\bibfnamefont {Ya.~B.}\ \bibnamefont {Zel'dovich}}\ and\ \bibinfo {author} {\bibfnamefont {I.~D.}\ \bibnamefont {Novikov}},\ }\bibfield  {title} {\enquote {\bibinfo {title} {The {Hypothesis} of {Cores} {Retarded} during {Expansion} and the {Hot} {Cosmological} {Model}},}\ }\href@noop {} {\bibfield  {journal} {\bibinfo  {journal} {Astronomicheskii Zhurnal}\ }\textbf {\bibinfo {volume} {43}},\ \bibinfo {pages} {758} (\bibinfo {year} {1966})}\BibitemShut {NoStop}%
\bibitem [{\citenamefont {Hawking}(1971)}]{hawking_gravitationally_1971}%
  \BibitemOpen
  \bibfield  {author} {\bibinfo {author} {\bibfnamefont {Stephen}\ \bibnamefont {Hawking}},\ }\bibfield  {title} {\enquote {\bibinfo {title} {Gravitationally {Collapsed} {Objects} of {Very} {Low} {Mass}},}\ }\href {\doibase 10.1093/mnras/152.1.75} {\bibfield  {journal} {\bibinfo  {journal} {Mon.~Not.~Roy.~Astron.~Soc.}\ }\textbf {\bibinfo {volume} {152}},\ \bibinfo {pages} {75--78} (\bibinfo {year} {1971})}\BibitemShut {NoStop}%
\bibitem [{\citenamefont {Carr}\ and\ \citenamefont {Hawking}(1974)}]{carr_black_1974}%
  \BibitemOpen
  \bibfield  {author} {\bibinfo {author} {\bibfnamefont {B.~J.}\ \bibnamefont {Carr}}\ and\ \bibinfo {author} {\bibfnamefont {S.~W.}\ \bibnamefont {Hawking}},\ }\bibfield  {title} {\enquote {\bibinfo {title} {Black {Holes} in the {Early} {Universe}},}\ }\href {\doibase 10.1093/mnras/168.2.399} {\bibfield  {journal} {\bibinfo  {journal} {Mon.~Not.~Roy.~Astron.~Soc.}\ }\textbf {\bibinfo {volume} {168}},\ \bibinfo {pages} {399--415} (\bibinfo {year} {1974})}\BibitemShut {NoStop}%
\bibitem [{\citenamefont {Khlopov}(2010)}]{Khlopov:2008qy}%
  \BibitemOpen
  \bibfield  {author} {\bibinfo {author} {\bibfnamefont {Maxim~Yu.}\ \bibnamefont {Khlopov}},\ }\bibfield  {title} {\enquote {\bibinfo {title} {{Primordial Black Holes}},}\ }\href {\doibase 10.1088/1674-4527/10/6/001} {\bibfield  {journal} {\bibinfo  {journal} {Res. Astron. Astrophys.}\ }\textbf {\bibinfo {volume} {10}},\ \bibinfo {pages} {495--528} (\bibinfo {year} {2010})},\ \Eprint {http://arxiv.org/abs/0801.0116} {arXiv:0801.0116 [astro-ph]} \BibitemShut {NoStop}%
\bibitem [{\citenamefont {Carr}\ and\ \citenamefont {Kühnel}(2020)}]{carr_primordial_2020}%
  \BibitemOpen
  \bibfield  {author} {\bibinfo {author} {\bibfnamefont {Bernard}\ \bibnamefont {Carr}}\ and\ \bibinfo {author} {\bibfnamefont {Florian}\ \bibnamefont {Kühnel}},\ }\bibfield  {title} {\enquote {\bibinfo {title} {Primordial {Black} {Holes} as {Dark} {Matter}: {Recent} {Developments}},}\ }\href {\doibase 10.1146/annurev-nucl-050520-125911} {\bibfield  {journal} {\bibinfo  {journal} {Annual Review of Nuclear and Particle Science}\ }\textbf {\bibinfo {volume} {70}},\ \bibinfo {pages} {355--394} (\bibinfo {year} {2020})},\ \Eprint {http://arxiv.org/abs/2006.02838} {arXiv:2006.02838 [astro-ph.CO]} \BibitemShut {NoStop}%
\bibitem [{\citenamefont {Carr}\ \emph {et~al.}(2021)\citenamefont {Carr}, \citenamefont {Kohri}, \citenamefont {Sendouda},\ and\ \citenamefont {Yokoyama}}]{carr_constraints_2021}%
  \BibitemOpen
  \bibfield  {author} {\bibinfo {author} {\bibfnamefont {Bernard}\ \bibnamefont {Carr}}, \bibinfo {author} {\bibfnamefont {Kazunori}\ \bibnamefont {Kohri}}, \bibinfo {author} {\bibfnamefont {Yuuiti}\ \bibnamefont {Sendouda}}, \ and\ \bibinfo {author} {\bibfnamefont {Jun’ichi}\ \bibnamefont {Yokoyama}},\ }\bibfield  {title} {\enquote {\bibinfo {title} {Constraints on primordial black holes},}\ }\href {\doibase 10.1088/1361-6633/ac1e31} {\bibfield  {journal} {\bibinfo  {journal} {Rep. Prog. Phys.}\ }\textbf {\bibinfo {volume} {84}},\ \bibinfo {pages} {116902} (\bibinfo {year} {2021})},\ \Eprint {http://arxiv.org/abs/2002.12778} {arXiv:2002.12778 [astro-ph.CO]} \BibitemShut {NoStop}%
\bibitem [{\citenamefont {Carr}\ and\ \citenamefont {Kuhnel}(2022)}]{carr_primordial_2022}%
  \BibitemOpen
  \bibfield  {author} {\bibinfo {author} {\bibfnamefont {Bernard}\ \bibnamefont {Carr}}\ and\ \bibinfo {author} {\bibfnamefont {Florian}\ \bibnamefont {Kuhnel}},\ }\bibfield  {title} {\enquote {\bibinfo {title} {{Primordial black holes as dark matter candidates}},}\ }\href {\doibase 10.21468/SciPostPhysLectNotes.48} {\bibfield  {journal} {\bibinfo  {journal} {SciPost Phys. Lect. Notes}\ }\textbf {\bibinfo {volume} {48}},\ \bibinfo {pages} {1} (\bibinfo {year} {2022})},\ \Eprint {http://arxiv.org/abs/2110.02821} {arXiv:2110.02821 [astro-ph.CO]} \BibitemShut {NoStop}%
\bibitem [{\citenamefont {Green}\ and\ \citenamefont {Kavanagh}(2021)}]{green_primordial_2021}%
  \BibitemOpen
  \bibfield  {author} {\bibinfo {author} {\bibfnamefont {Anne~M.}\ \bibnamefont {Green}}\ and\ \bibinfo {author} {\bibfnamefont {Bradley~J.}\ \bibnamefont {Kavanagh}},\ }\bibfield  {title} {\enquote {\bibinfo {title} {{Primordial Black Holes as a dark matter candidate}},}\ }\href {\doibase 10.1088/1361-6471/abc534} {\bibfield  {journal} {\bibinfo  {journal} {J. Phys. G}\ }\textbf {\bibinfo {volume} {48}},\ \bibinfo {pages} {043001} (\bibinfo {year} {2021})},\ \Eprint {http://arxiv.org/abs/2007.10722} {arXiv:2007.10722 [astro-ph.CO]} \BibitemShut {NoStop}%
\bibitem [{\citenamefont {Kim}(2021)}]{Kim:2020ngi}%
  \BibitemOpen
  \bibfield  {author} {\bibinfo {author} {\bibfnamefont {Hyungjin}\ \bibnamefont {Kim}},\ }\bibfield  {title} {\enquote {\bibinfo {title} {{A constraint on light primordial black holes from the interstellar medium temperature}},}\ }\href {\doibase 10.1093/mnras/stab1222} {\bibfield  {journal} {\bibinfo  {journal} {Mon. Not. Roy. Astron. Soc.}\ }\textbf {\bibinfo {volume} {504}},\ \bibinfo {pages} {5475--5484} (\bibinfo {year} {2021})},\ \Eprint {http://arxiv.org/abs/2007.07739} {arXiv:2007.07739 [hep-ph]} \BibitemShut {NoStop}%
\bibitem [{\citenamefont {Berteaud}\ \emph {et~al.}(2022)\citenamefont {Berteaud}, \citenamefont {Calore}, \citenamefont {Iguaz}, \citenamefont {Serpico},\ and\ \citenamefont {Siegert}}]{Berteaud:2022tws}%
  \BibitemOpen
  \bibfield  {author} {\bibinfo {author} {\bibfnamefont {J.}~\bibnamefont {Berteaud}}, \bibinfo {author} {\bibfnamefont {F.}~\bibnamefont {Calore}}, \bibinfo {author} {\bibfnamefont {J.}~\bibnamefont {Iguaz}}, \bibinfo {author} {\bibfnamefont {P.~D.}\ \bibnamefont {Serpico}}, \ and\ \bibinfo {author} {\bibfnamefont {T.}~\bibnamefont {Siegert}},\ }\bibfield  {title} {\enquote {\bibinfo {title} {{Strong constraints on primordial black hole dark matter from 16~years of INTEGRAL/SPI observations}},}\ }\href {\doibase 10.1103/PhysRevD.106.023030} {\bibfield  {journal} {\bibinfo  {journal} {Phys. Rev. D}\ }\textbf {\bibinfo {volume} {106}},\ \bibinfo {pages} {023030} (\bibinfo {year} {2022})},\ \Eprint {http://arxiv.org/abs/2202.07483} {arXiv:2202.07483 [astro-ph.HE]} \BibitemShut {NoStop}%
\bibitem [{\citenamefont {Carr}\ \emph {et~al.}(2024)\citenamefont {Carr}, \citenamefont {Clesse}, \citenamefont {Garcia-Bellido}, \citenamefont {Hawkins},\ and\ \citenamefont {Kuhnel}}]{carr_observational_2024}%
  \BibitemOpen
  \bibfield  {author} {\bibinfo {author} {\bibfnamefont {Bernard}\ \bibnamefont {Carr}}, \bibinfo {author} {\bibfnamefont {Sebastien}\ \bibnamefont {Clesse}}, \bibinfo {author} {\bibfnamefont {Juan}\ \bibnamefont {Garcia-Bellido}}, \bibinfo {author} {\bibfnamefont {Michael}\ \bibnamefont {Hawkins}}, \ and\ \bibinfo {author} {\bibfnamefont {Florian}\ \bibnamefont {Kuhnel}},\ }\bibfield  {title} {\enquote {\bibinfo {title} {{Observational evidence for primordial black holes: A positivist perspective}},}\ }\href {\doibase 10.1016/j.physrep.2023.11.005} {\bibfield  {journal} {\bibinfo  {journal} {Phys. Rept.}\ }\textbf {\bibinfo {volume} {1054}},\ \bibinfo {pages} {1--68} (\bibinfo {year} {2024})},\ \Eprint {http://arxiv.org/abs/2306.03903} {arXiv:2306.03903 [astro-ph.CO]} \BibitemShut {NoStop}%
\bibitem [{\citenamefont {Escrivà}\ \emph {et~al.}(2024)\citenamefont {Escrivà}, \citenamefont {Kühnel},\ and\ \citenamefont {Tada}}]{escriva_primordial_2024}%
  \BibitemOpen
  \bibfield  {author} {\bibinfo {author} {\bibfnamefont {Albert}\ \bibnamefont {Escrivà}}, \bibinfo {author} {\bibfnamefont {Florian}\ \bibnamefont {Kühnel}}, \ and\ \bibinfo {author} {\bibfnamefont {Yuichiro}\ \bibnamefont {Tada}},\ }\bibfield  {title} {\enquote {\bibinfo {title} {Chapter 4 - primordial black holes},}\ }in\ \href {\doibase https://doi.org/10.1016/B978-0-32-395636-9.00012-8} {\emph {\bibinfo {booktitle} {Black Holes in the Era of Gravitational-Wave Astronomy}}},\ \bibinfo {editor} {edited by\ \bibinfo {editor} {\bibfnamefont {Manuel~Arca}\ \bibnamefont {Sedda}}, \bibinfo {editor} {\bibfnamefont {Elisa}\ \bibnamefont {Bortolas}}, \ and\ \bibinfo {editor} {\bibfnamefont {Mario}\ \bibnamefont {Spera}}}\ (\bibinfo  {publisher} {Elsevier},\ \bibinfo {year} {2024})\ pp.\ \bibinfo {pages} {261--377},\ \Eprint {http://arxiv.org/abs/2211.05767} {arXiv:2211.05767 [astro-ph.CO]} \BibitemShut {NoStop}%
\bibitem [{\citenamefont {Gorton}\ and\ \citenamefont {Green}(2024)}]{gorton_how_2024}%
  \BibitemOpen
  \bibfield  {author} {\bibinfo {author} {\bibfnamefont {Matthew}\ \bibnamefont {Gorton}}\ and\ \bibinfo {author} {\bibfnamefont {Anne~M.}\ \bibnamefont {Green}},\ }\bibfield  {title} {\enquote {\bibinfo {title} {{How open is the asteroid-mass primordial black hole window?}}}\ }\href {\doibase 10.21468/SciPostPhys.17.2.032} {\bibfield  {journal} {\bibinfo  {journal} {SciPost Phys.}\ }\textbf {\bibinfo {volume} {17}},\ \bibinfo {pages} {032} (\bibinfo {year} {2024})},\ \Eprint {http://arxiv.org/abs/2403.03839} {arXiv:2403.03839 [astro-ph.CO]} \BibitemShut {NoStop}%
\bibitem [{\citenamefont {De~la Torre~Luque}\ \emph {et~al.}(2024)\citenamefont {De~la Torre~Luque}, \citenamefont {Koechler},\ and\ \citenamefont {Balaji}}]{DelaTorreLuque:2024qms}%
  \BibitemOpen
  \bibfield  {author} {\bibinfo {author} {\bibfnamefont {Pedro}\ \bibnamefont {De~la Torre~Luque}}, \bibinfo {author} {\bibfnamefont {Jordan}\ \bibnamefont {Koechler}}, \ and\ \bibinfo {author} {\bibfnamefont {Shyam}\ \bibnamefont {Balaji}},\ }\bibfield  {title} {\enquote {\bibinfo {title} {{Refining Galactic primordial black hole evaporation constraints}},}\ }\href {\doibase 10.1103/PhysRevD.110.123022} {\bibfield  {journal} {\bibinfo  {journal} {Phys. Rev. D}\ }\textbf {\bibinfo {volume} {110}},\ \bibinfo {pages} {123022} (\bibinfo {year} {2024})},\ \Eprint {http://arxiv.org/abs/2406.11949} {arXiv:2406.11949 [astro-ph.HE]} \BibitemShut {NoStop}%
\bibitem [{\citenamefont {Hawking}(1974)}]{hawking_black_1974}%
  \BibitemOpen
  \bibfield  {author} {\bibinfo {author} {\bibfnamefont {S.~W.}\ \bibnamefont {Hawking}},\ }\bibfield  {title} {\enquote {\bibinfo {title} {Black hole explosions?}}\ }\href {\doibase 10.1038/248030a0} {\bibfield  {journal} {\bibinfo  {journal} {Nature}\ }\textbf {\bibinfo {volume} {248}},\ \bibinfo {pages} {30--31} (\bibinfo {year} {1974})}\BibitemShut {NoStop}%
\bibitem [{\citenamefont {Hawking}(1975)}]{hawking_particle_1975}%
  \BibitemOpen
  \bibfield  {author} {\bibinfo {author} {\bibfnamefont {S.~W.}\ \bibnamefont {Hawking}},\ }\bibfield  {title} {\enquote {\bibinfo {title} {Particle creation by black holes},}\ }\href {\doibase 10.1007/BF02345020} {\bibfield  {journal} {\bibinfo  {journal} {Commun.~Math.~Phys.}\ }\textbf {\bibinfo {volume} {43}},\ \bibinfo {pages} {199--220} (\bibinfo {year} {1975})}\BibitemShut {NoStop}%
\bibitem [{\citenamefont {Page}(1976{\natexlab{a}})}]{page_particle_1976}%
  \BibitemOpen
  \bibfield  {author} {\bibinfo {author} {\bibfnamefont {Don~N.}\ \bibnamefont {Page}},\ }\bibfield  {title} {\enquote {\bibinfo {title} {Particle emission rates from a black hole: {Massless} particles from an uncharged, nonrotating hole},}\ }\href {\doibase 10.1103/PhysRevD.13.198} {\bibfield  {journal} {\bibinfo  {journal} {Phys.~Rev.~D}\ }\textbf {\bibinfo {volume} {13}},\ \bibinfo {pages} {198--206} (\bibinfo {year} {1976}{\natexlab{a}})}\BibitemShut {NoStop}%
\bibitem [{\citenamefont {Page}(1976{\natexlab{b}})}]{page_particle_1976-1}%
  \BibitemOpen
  \bibfield  {author} {\bibinfo {author} {\bibfnamefont {Don~N.}\ \bibnamefont {Page}},\ }\bibfield  {title} {\enquote {\bibinfo {title} {Particle emission rates from a black hole. {II}. {Massless} particles from a rotating hole},}\ }\href {\doibase 10.1103/PhysRevD.14.3260} {\bibfield  {journal} {\bibinfo  {journal} {Phys.~Rev.~D}\ }\textbf {\bibinfo {volume} {14}},\ \bibinfo {pages} {3260--3273} (\bibinfo {year} {1976}{\natexlab{b}})}\BibitemShut {NoStop}%
\bibitem [{\citenamefont {Page}(1977)}]{page_particle_1977}%
  \BibitemOpen
  \bibfield  {author} {\bibinfo {author} {\bibfnamefont {Don~N.}\ \bibnamefont {Page}},\ }\bibfield  {title} {\enquote {\bibinfo {title} {Particle emission rates from a black hole. {III}. {Charged} leptons from a nonrotating hole},}\ }\href {\doibase 10.1103/PhysRevD.16.2402} {\bibfield  {journal} {\bibinfo  {journal} {Phys.~Rev.~D}\ }\textbf {\bibinfo {volume} {16}},\ \bibinfo {pages} {2402--2411} (\bibinfo {year} {1977})}\BibitemShut {NoStop}%
\bibitem [{\citenamefont {MacGibbon}\ and\ \citenamefont {Webber}(1990)}]{macgibbon_quark-_1990}%
  \BibitemOpen
  \bibfield  {author} {\bibinfo {author} {\bibfnamefont {Jane~H.}\ \bibnamefont {MacGibbon}}\ and\ \bibinfo {author} {\bibfnamefont {B.~R.}\ \bibnamefont {Webber}},\ }\bibfield  {title} {\enquote {\bibinfo {title} {Quark- and gluon-jet emission from primordial black holes: {The} instantaneous spectra},}\ }\href {\doibase 10.1103/PhysRevD.41.3052} {\bibfield  {journal} {\bibinfo  {journal} {Phys. Rev. D}\ }\textbf {\bibinfo {volume} {41}},\ \bibinfo {pages} {3052--3079} (\bibinfo {year} {1990})}\BibitemShut {NoStop}%
\bibitem [{\citenamefont {MacGibbon}(1991)}]{macgibbon_quark-_1991}%
  \BibitemOpen
  \bibfield  {author} {\bibinfo {author} {\bibfnamefont {Jane~H.}\ \bibnamefont {MacGibbon}},\ }\bibfield  {title} {\enquote {\bibinfo {title} {Quark- and gluon-jet emission from primordial black holes. {II}. {The} emission over the black-hole lifetime},}\ }\href {\doibase 10.1103/PhysRevD.44.376} {\bibfield  {journal} {\bibinfo  {journal} {Phys.~Rev.~D}\ }\textbf {\bibinfo {volume} {44}},\ \bibinfo {pages} {376--392} (\bibinfo {year} {1991})}\BibitemShut {NoStop}%
\bibitem [{\citenamefont {Glicenstein}\ \emph {et~al.}(2013)\citenamefont {Glicenstein}, \citenamefont {Barnacka}, \citenamefont {Vivier},\ and\ \citenamefont {Herr}}]{glicenstein_limits_2013}%
  \BibitemOpen
  \bibfield  {author} {\bibinfo {author} {\bibfnamefont {J-F.}\ \bibnamefont {Glicenstein}}, \bibinfo {author} {\bibfnamefont {A.}~\bibnamefont {Barnacka}}, \bibinfo {author} {\bibfnamefont {M.}~\bibnamefont {Vivier}}, \ and\ \bibinfo {author} {\bibfnamefont {T.}~\bibnamefont {Herr}} (\bibinfo {collaboration} {H.E.S.S.}),\ }\bibfield  {title} {\enquote {\bibinfo {title} {{Limits on Primordial Black Hole evaporation with the H.E.S.S. array of Cherenkov telescopes}},}\ }in\ \href@noop {} {\emph {\bibinfo {booktitle} {{33rd International Cosmic Ray Conference}}}}\ (\bibinfo {year} {2013})\ p.\ \bibinfo {pages} {0930},\ \Eprint {http://arxiv.org/abs/1307.4898} {arXiv:1307.4898 [astro-ph.HE]} \BibitemShut {NoStop}%
\bibitem [{\citenamefont {Abdo}\ \emph {et~al.}(2015)\citenamefont {Abdo} \emph {et~al.}}]{abdo_milagro_2015}%
  \BibitemOpen
  \bibfield  {author} {\bibinfo {author} {\bibfnamefont {A.~A.}\ \bibnamefont {Abdo}} \emph {et~al.},\ }\bibfield  {title} {\enquote {\bibinfo {title} {{Milagro Limits and HAWC Sensitivity for the Rate-Density of Evaporating Primordial Black Holes}},}\ }\href {\doibase 10.1016/j.astropartphys.2014.10.007} {\bibfield  {journal} {\bibinfo  {journal} {Astropart. Phys.}\ }\textbf {\bibinfo {volume} {64}},\ \bibinfo {pages} {4--12} (\bibinfo {year} {2015})},\ \Eprint {http://arxiv.org/abs/1407.1686} {arXiv:1407.1686 [astro-ph.HE]} \BibitemShut {NoStop}%
\bibitem [{\citenamefont {Archambault}(2018)}]{archambault_search_2017}%
  \BibitemOpen
  \bibfield  {author} {\bibinfo {author} {\bibfnamefont {Simon}\ \bibnamefont {Archambault}} (\bibinfo {collaboration} {VERITAS}),\ }\bibfield  {title} {\enquote {\bibinfo {title} {{Search for Primordial Black Hole Evaporation with VERITAS}},}\ }\href {\doibase 10.22323/1.301.0691} {\bibfield  {journal} {\bibinfo  {journal} {PoS}\ }\textbf {\bibinfo {volume} {ICRC2017}},\ \bibinfo {pages} {691} (\bibinfo {year} {2018})},\ \Eprint {http://arxiv.org/abs/1709.00307} {arXiv:1709.00307 [astro-ph.HE]} \BibitemShut {NoStop}%
\bibitem [{\citenamefont {Ackermann}\ \emph {et~al.}(2018)\citenamefont {Ackermann} \emph {et~al.}}]{the_fermi-lat_collaboration_search_2018}%
  \BibitemOpen
  \bibfield  {author} {\bibinfo {author} {\bibfnamefont {M.}~\bibnamefont {Ackermann}} \emph {et~al.} (\bibinfo {collaboration} {Fermi-LAT}),\ }\bibfield  {title} {\enquote {\bibinfo {title} {{Search for Gamma-Ray Emission from Local Primordial Black Holes with the Fermi Large Area Telescope}},}\ }\href {\doibase 10.3847/1538-4357/aaac7b} {\bibfield  {journal} {\bibinfo  {journal} {Astrophys. J.}\ }\textbf {\bibinfo {volume} {857}},\ \bibinfo {pages} {49} (\bibinfo {year} {2018})},\ \Eprint {http://arxiv.org/abs/1802.00100} {arXiv:1802.00100 [astro-ph.HE]} \BibitemShut {NoStop}%
\bibitem [{\citenamefont {Albert}\ \emph {et~al.}(2020)\citenamefont {Albert} \emph {et~al.}}]{albert_constraining_2020}%
  \BibitemOpen
  \bibfield  {author} {\bibinfo {author} {\bibfnamefont {A.}~\bibnamefont {Albert}} \emph {et~al.} (\bibinfo {collaboration} {HAWC}),\ }\bibfield  {title} {\enquote {\bibinfo {title} {{Constraining the Local Burst Rate Density of Primordial Black Holes with HAWC}},}\ }\href {\doibase 10.1088/1475-7516/2020/04/026} {\bibfield  {journal} {\bibinfo  {journal} {JCAP}\ }\textbf {\bibinfo {volume} {04}},\ \bibinfo {pages} {026} (\bibinfo {year} {2020})},\ \Eprint {http://arxiv.org/abs/1911.04356} {arXiv:1911.04356 [astro-ph.HE]} \BibitemShut {NoStop}%
\bibitem [{\citenamefont {Chiba}\ and\ \citenamefont {Yokoyama}(2017)}]{chiba_spin_2017}%
  \BibitemOpen
  \bibfield  {author} {\bibinfo {author} {\bibfnamefont {Takeshi}\ \bibnamefont {Chiba}}\ and\ \bibinfo {author} {\bibfnamefont {Shuichiro}\ \bibnamefont {Yokoyama}},\ }\bibfield  {title} {\enquote {\bibinfo {title} {{Spin Distribution of Primordial Black Holes}},}\ }\href {\doibase 10.1093/ptep/ptx087} {\bibfield  {journal} {\bibinfo  {journal} {PTEP}\ }\textbf {\bibinfo {volume} {2017}},\ \bibinfo {pages} {083E01} (\bibinfo {year} {2017})},\ \Eprint {http://arxiv.org/abs/1704.06573} {arXiv:1704.06573 [gr-qc]} \BibitemShut {NoStop}%
\bibitem [{\citenamefont {De~Luca}\ \emph {et~al.}(2019)\citenamefont {De~Luca}, \citenamefont {Desjacques}, \citenamefont {Franciolini}, \citenamefont {Malhotra},\ and\ \citenamefont {Riotto}}]{de_luca_initial_2019}%
  \BibitemOpen
  \bibfield  {author} {\bibinfo {author} {\bibfnamefont {V.}~\bibnamefont {De~Luca}}, \bibinfo {author} {\bibfnamefont {V.}~\bibnamefont {Desjacques}}, \bibinfo {author} {\bibfnamefont {G.}~\bibnamefont {Franciolini}}, \bibinfo {author} {\bibfnamefont {A.}~\bibnamefont {Malhotra}}, \ and\ \bibinfo {author} {\bibfnamefont {A.}~\bibnamefont {Riotto}},\ }\bibfield  {title} {\enquote {\bibinfo {title} {{The initial spin probability distribution of primordial black holes}},}\ }\href {\doibase 10.1088/1475-7516/2019/05/018} {\bibfield  {journal} {\bibinfo  {journal} {JCAP}\ }\textbf {\bibinfo {volume} {05}},\ \bibinfo {pages} {018} (\bibinfo {year} {2019})},\ \Eprint {http://arxiv.org/abs/1903.01179} {arXiv:1903.01179 [astro-ph.CO]} \BibitemShut {NoStop}%
\bibitem [{\citenamefont {De~Luca}\ \emph {et~al.}(2020{\natexlab{a}})\citenamefont {De~Luca}, \citenamefont {Franciolini}, \citenamefont {Pani},\ and\ \citenamefont {Riotto}}]{de_luca_evolution_2020}%
  \BibitemOpen
  \bibfield  {author} {\bibinfo {author} {\bibfnamefont {V.}~\bibnamefont {De~Luca}}, \bibinfo {author} {\bibfnamefont {G.}~\bibnamefont {Franciolini}}, \bibinfo {author} {\bibfnamefont {P.}~\bibnamefont {Pani}}, \ and\ \bibinfo {author} {\bibfnamefont {A.}~\bibnamefont {Riotto}},\ }\bibfield  {title} {\enquote {\bibinfo {title} {{The evolution of primordial black holes and their final observable spins}},}\ }\href {\doibase 10.1088/1475-7516/2020/04/052} {\bibfield  {journal} {\bibinfo  {journal} {JCAP}\ }\textbf {\bibinfo {volume} {04}},\ \bibinfo {pages} {052} (\bibinfo {year} {2020}{\natexlab{a}})},\ \Eprint {http://arxiv.org/abs/2003.02778} {arXiv:2003.02778 [astro-ph.CO]} \BibitemShut {NoStop}%
\bibitem [{\citenamefont {Chongchitnan}\ and\ \citenamefont {Silk}(2021)}]{chongchitnan_extreme-value_2021}%
  \BibitemOpen
  \bibfield  {author} {\bibinfo {author} {\bibfnamefont {Siri}\ \bibnamefont {Chongchitnan}}\ and\ \bibinfo {author} {\bibfnamefont {Joseph}\ \bibnamefont {Silk}},\ }\bibfield  {title} {\enquote {\bibinfo {title} {{Extreme-value statistics of the spin of primordial black holes}},}\ }\href {\doibase 10.1103/PhysRevD.104.083018} {\bibfield  {journal} {\bibinfo  {journal} {Phys. Rev. D}\ }\textbf {\bibinfo {volume} {104}},\ \bibinfo {pages} {083018} (\bibinfo {year} {2021})},\ \Eprint {http://arxiv.org/abs/2109.12268} {arXiv:2109.12268 [astro-ph.CO]} \BibitemShut {NoStop}%
\bibitem [{\citenamefont {Alonso-Monsalve}\ and\ \citenamefont {Kaiser}(2024)}]{alonso-monsalve_primordial_2024}%
  \BibitemOpen
  \bibfield  {author} {\bibinfo {author} {\bibfnamefont {Elba}\ \bibnamefont {Alonso-Monsalve}}\ and\ \bibinfo {author} {\bibfnamefont {David~I.}\ \bibnamefont {Kaiser}},\ }\bibfield  {title} {\enquote {\bibinfo {title} {{Primordial Black Holes with QCD Color Charge}},}\ }\href {\doibase 10.1103/PhysRevLett.132.231402} {\bibfield  {journal} {\bibinfo  {journal} {Phys. Rev. Lett.}\ }\textbf {\bibinfo {volume} {132}},\ \bibinfo {pages} {231402} (\bibinfo {year} {2024})},\ \Eprint {http://arxiv.org/abs/2310.16877} {arXiv:2310.16877 [hep-ph]} \BibitemShut {NoStop}%
\bibitem [{Cha()}]{ChargeSpinNote}%
  \BibitemOpen
  \href@noop {} {}\bibinfo {note} {In general, PBHs that form with significant spin or charge have lower surface gravity than PBHs that form with the same mass but vanishing spin and charge. The lower surface gravity yields a lower Hawking temperature. However, Page demonstrated that an initially rotating black hole will preferentially emit particles with spin aligned with the black hole's angular momentum, causing it to ``spin down'' to within $1\%$ of the emission rate of a same-mass Schwarzschild black hole before half of its energy is radiated away. The exploding PBHs we consider have present-day masses $M\leq 10^8 \, {\rm g}$ and must have formed with initial mass $M_i\sim10^{14} \, {\rm g}$. Thus, these PBHs have radiated approximately all of their formation mass and Page's statement should apply, leading us to assume that present-day exploding PBHs have no spin. He further demonstrated that an initially maximally rotating PBH will have a lifetime that differs from that of its Schwarzschild counterpart
  only by a factor of $2.32$~\cite{page_particle_1976-1}, so we neglect initial spin here. Meanwhile, PBHs discharge any initial charge even more efficiently, leaving stochastic charge fluctuations of order ${\cal O} (q) \ll \sqrt{G} \, M$ \cite{gibbons_vacuum_1975,carter_charge_1974}, so we likewise neglect initial charge for the PBHs.}\BibitemShut {Stop}%
\bibitem [{\citenamefont {Dvali}(2018)}]{Dvali:2018xpy}%
  \BibitemOpen
  \bibfield  {author} {\bibinfo {author} {\bibfnamefont {Gia}\ \bibnamefont {Dvali}},\ }\bibfield  {title} {\enquote {\bibinfo {title} {{A Microscopic Model of Holography: Survival by the Burden of Memory}},}\ }\href@noop {} {\  (\bibinfo {year} {2018})},\ \Eprint {http://arxiv.org/abs/1810.02336} {arXiv:1810.02336 [hep-th]} \BibitemShut {NoStop}%
\bibitem [{\citenamefont {Dvali}\ \emph {et~al.}(2020)\citenamefont {Dvali}, \citenamefont {Eisemann}, \citenamefont {Michel},\ and\ \citenamefont {Zell}}]{Dvali:2020wft}%
  \BibitemOpen
  \bibfield  {author} {\bibinfo {author} {\bibfnamefont {Gia}\ \bibnamefont {Dvali}}, \bibinfo {author} {\bibfnamefont {Lukas}\ \bibnamefont {Eisemann}}, \bibinfo {author} {\bibfnamefont {Marco}\ \bibnamefont {Michel}}, \ and\ \bibinfo {author} {\bibfnamefont {Sebastian}\ \bibnamefont {Zell}},\ }\bibfield  {title} {\enquote {\bibinfo {title} {{Black hole metamorphosis and stabilization by memory burden}},}\ }\href {\doibase 10.1103/PhysRevD.102.103523} {\bibfield  {journal} {\bibinfo  {journal} {Phys. Rev. D}\ }\textbf {\bibinfo {volume} {102}},\ \bibinfo {pages} {103523} (\bibinfo {year} {2020})},\ \Eprint {http://arxiv.org/abs/2006.00011} {arXiv:2006.00011 [hep-th]} \BibitemShut {NoStop}%
\bibitem [{\citenamefont {Zantedeschi}\ and\ \citenamefont {Visinelli}(2024)}]{zantedeschi_ultralight_2025}%
  \BibitemOpen
  \bibfield  {author} {\bibinfo {author} {\bibfnamefont {Michael}\ \bibnamefont {Zantedeschi}}\ and\ \bibinfo {author} {\bibfnamefont {Luca}\ \bibnamefont {Visinelli}},\ }\bibfield  {title} {\enquote {\bibinfo {title} {{Ultralight Black Holes as Sources of High-Energy Particles}},}\ }\href@noop {} {\  (\bibinfo {year} {2024})},\ \Eprint {http://arxiv.org/abs/2410.07037} {arXiv:2410.07037 [astro-ph.HE]} \BibitemShut {NoStop}%
\bibitem [{\citenamefont {Boccia}\ and\ \citenamefont {Iocco}(2025)}]{boccia_strike_2025}%
  \BibitemOpen
  \bibfield  {author} {\bibinfo {author} {\bibfnamefont {Andrea}\ \bibnamefont {Boccia}}\ and\ \bibinfo {author} {\bibfnamefont {Fabio}\ \bibnamefont {Iocco}},\ }\href@noop {} {\enquote {\bibinfo {title} {{A strike of luck: could the KM3-230213A event be caused by an evaporating primordial black hole?}}}\ } (\bibinfo {year} {2025}),\ \Eprint {http://arxiv.org/abs/2502.19245} {arXiv:2502.19245 [astro-ph.HE]} \BibitemShut {NoStop}%
\bibitem [{\citenamefont {Teukolsky}\ and\ \citenamefont {Press}(1974)}]{teukolsky_perturbations_1974}%
  \BibitemOpen
  \bibfield  {author} {\bibinfo {author} {\bibfnamefont {S.~A.}\ \bibnamefont {Teukolsky}}\ and\ \bibinfo {author} {\bibfnamefont {W.~H.}\ \bibnamefont {Press}},\ }\bibfield  {title} {\enquote {\bibinfo {title} {Perturbations of a rotating black hole. {III}. {Interaction} of the hole with gravitational and electromagnetic radiation.}}\ }\href {\doibase 10.1086/153180} {\bibfield  {journal} {\bibinfo  {journal} {Astrophys.~J.}\ }\textbf {\bibinfo {volume} {193}},\ \bibinfo {pages} {443--461} (\bibinfo {year} {1974})}\BibitemShut {NoStop}%
\bibitem [{\citenamefont {Teukolsky}(1973)}]{teukolsky_perturbations_1973}%
  \BibitemOpen
  \bibfield  {author} {\bibinfo {author} {\bibfnamefont {Saul~A.}\ \bibnamefont {Teukolsky}},\ }\bibfield  {title} {\enquote {\bibinfo {title} {Perturbations of a {Rotating} {Black} {Hole}. {I}. {Fundamental} {Equations} for {Gravitational}, {Electromagnetic}, and {Neutrino}-{Field} {Perturbations}},}\ }\href {\doibase 10.1086/152444} {\bibfield  {journal} {\bibinfo  {journal} {Astrophys.~J.}\ }\textbf {\bibinfo {volume} {185}},\ \bibinfo {pages} {635} (\bibinfo {year} {1973})}\BibitemShut {NoStop}%
\bibitem [{\citenamefont {Perez-Gonzalez}(2025)}]{Perez-Gonzalez:2025try}%
  \BibitemOpen
  \bibfield  {author} {\bibinfo {author} {\bibfnamefont {Yuber~F.}\ \bibnamefont {Perez-Gonzalez}},\ }\bibfield  {title} {\enquote {\bibinfo {title} {{Page Time of Primordial Black Holes in the Standard Model and Beyond}},}\ }\href@noop {} {\  (\bibinfo {year} {2025})},\ \Eprint {http://arxiv.org/abs/2502.04430} {arXiv:2502.04430 [astro-ph.CO]} \BibitemShut {NoStop}%
\bibitem [{BSM()}]{BSMnote}%
  \BibitemOpen
  \href@noop {} {}\bibinfo {note} {If elementary particles beyond those of the SM exist, they would also be included among the primary emission spectra during PBH Hawking evaporation. On the one hand, increasing the total number of degrees of freedom emitted from a PBH of a given mass would reduce the number of high-energy neutrinos emitted during primary emission for small-mass PBHs near the end of their lifetimes. On the other hand, depending on whether the new particles couple to neutrinos, their presence could increase the number of high-energy neutrinos emitted via secondary emission. Such effects remain strongly model-dependent. See, e.g., Refs.~\cite{Wu:2024uxa,Khlopov:2024nqp,Choi:2025hqt,Baker:2025cff,Olinto:2025elp}. As a concrete example, adding a scalar field with $m \geq 100 \, {\rm GeV}$ would increase the Page factor $f(M)$ for $M\lesssim 10^{11} \, {\rm g}$. Adding an additional heavy degree of freedom would not modify the primary neutrino spectrum Eq.~(\ref{eqn:PrimarySpectraSchwarzschild}),
  because this only depends on the number of neutrino DOFs. The increase in the Page factor results in an increase in the overall mass-loss rate in Eq.~(\ref{eqn:MassEvolution}) and thus a decrease in the integrated total neutrino emission rate in Eq.~(\ref{Nnutotal}). We find that for this case of a single spin-0 dark DOF with $m = 100\, {\rm GeV}$, the value of $N_{\nu}$ decreases by $\sim 2\%$ for $Q_{\rm min} \gtrsim 100 \, {\rm GeV}$.}\BibitemShut {Stop}%
\bibitem [{\citenamefont {Arbey}\ and\ \citenamefont {Auffinger}(2021)}]{arbey_physics_2021}%
  \BibitemOpen
  \bibfield  {author} {\bibinfo {author} {\bibfnamefont {Alexandre}\ \bibnamefont {Arbey}}\ and\ \bibinfo {author} {\bibfnamefont {J\'er\'emy}\ \bibnamefont {Auffinger}},\ }\bibfield  {title} {\enquote {\bibinfo {title} {{Physics Beyond the Standard Model with BlackHawk v2.0}},}\ }\href {\doibase 10.1140/epjc/s10052-021-09702-8} {\bibfield  {journal} {\bibinfo  {journal} {Eur. Phys. J. C}\ }\textbf {\bibinfo {volume} {81}},\ \bibinfo {pages} {910} (\bibinfo {year} {2021})},\ \Eprint {http://arxiv.org/abs/2108.02737} {arXiv:2108.02737 [gr-qc]} \BibitemShut {NoStop}%
\bibitem [{\citenamefont {Arbey}\ and\ \citenamefont {Auffinger}(2019)}]{arbey_blackhawk_2019}%
  \BibitemOpen
  \bibfield  {author} {\bibinfo {author} {\bibfnamefont {Alexandre}\ \bibnamefont {Arbey}}\ and\ \bibinfo {author} {\bibfnamefont {J\'er\'emy}\ \bibnamefont {Auffinger}},\ }\bibfield  {title} {\enquote {\bibinfo {title} {{BlackHawk: A public code for calculating the Hawking evaporation spectra of any black hole distribution}},}\ }\href {\doibase 10.1140/epjc/s10052-019-7161-1} {\bibfield  {journal} {\bibinfo  {journal} {Eur. Phys. J. C}\ }\textbf {\bibinfo {volume} {79}},\ \bibinfo {pages} {693} (\bibinfo {year} {2019})},\ \Eprint {http://arxiv.org/abs/1905.04268} {arXiv:1905.04268 [gr-qc]} \BibitemShut {NoStop}%
\bibitem [{\citenamefont {Bauer}\ \emph {et~al.}(2021)\citenamefont {Bauer}, \citenamefont {Rodd},\ and\ \citenamefont {Webber}}]{Bauer:2020jay}%
  \BibitemOpen
  \bibfield  {author} {\bibinfo {author} {\bibfnamefont {Christian~W.}\ \bibnamefont {Bauer}}, \bibinfo {author} {\bibfnamefont {Nicholas~L.}\ \bibnamefont {Rodd}}, \ and\ \bibinfo {author} {\bibfnamefont {Bryan~R.}\ \bibnamefont {Webber}},\ }\bibfield  {title} {\enquote {\bibinfo {title} {{Dark matter spectra from the electroweak to the Planck scale}},}\ }\href {\doibase 10.1007/JHEP06(2021)121} {\bibfield  {journal} {\bibinfo  {journal} {JHEP}\ }\textbf {\bibinfo {volume} {06}},\ \bibinfo {pages} {121} (\bibinfo {year} {2021})},\ \Eprint {http://arxiv.org/abs/2007.15001} {arXiv:2007.15001 [hep-ph]} \BibitemShut {NoStop}%
\bibitem [{\citenamefont {Gundlach}\ and\ \citenamefont {Martin-Garcia}(2007)}]{gundlach_critical_2007}%
  \BibitemOpen
  \bibfield  {author} {\bibinfo {author} {\bibfnamefont {Carsten}\ \bibnamefont {Gundlach}}\ and\ \bibinfo {author} {\bibfnamefont {Jose~M.}\ \bibnamefont {Martin-Garcia}},\ }\bibfield  {title} {\enquote {\bibinfo {title} {{Critical phenomena in gravitational collapse}},}\ }\href {\doibase 10.12942/lrr-2007-5} {\bibfield  {journal} {\bibinfo  {journal} {Living Rev. Rel.}\ }\textbf {\bibinfo {volume} {10}},\ \bibinfo {pages} {5} (\bibinfo {year} {2007})},\ \Eprint {http://arxiv.org/abs/0711.4620} {arXiv:0711.4620 [gr-qc]} \BibitemShut {NoStop}%
\bibitem [{\citenamefont {Escriv\`a}(2022)}]{Escriva:2021aeh}%
  \BibitemOpen
  \bibfield  {author} {\bibinfo {author} {\bibfnamefont {Albert}\ \bibnamefont {Escriv\`a}},\ }\bibfield  {title} {\enquote {\bibinfo {title} {{PBH Formation from Spherically Symmetric Hydrodynamical Perturbations: A Review}},}\ }\href {\doibase 10.3390/universe8020066} {\bibfield  {journal} {\bibinfo  {journal} {Universe}\ }\textbf {\bibinfo {volume} {8}},\ \bibinfo {pages} {66} (\bibinfo {year} {2022})},\ \Eprint {http://arxiv.org/abs/2111.12693} {arXiv:2111.12693 [gr-qc]} \BibitemShut {NoStop}%
\bibitem [{\citenamefont {Aghanim}\ \emph {et~al.}(2020)\citenamefont {Aghanim} \emph {et~al.}}]{planck_collaboration_planck_2020}%
  \BibitemOpen
  \bibfield  {author} {\bibinfo {author} {\bibfnamefont {N.}~\bibnamefont {Aghanim}} \emph {et~al.} (\bibinfo {collaboration} {Planck}),\ }\bibfield  {title} {\enquote {\bibinfo {title} {{Planck 2018 results. VI. Cosmological parameters}},}\ }\href {\doibase 10.1051/0004-6361/201833910} {\bibfield  {journal} {\bibinfo  {journal} {Astron. Astrophys.}\ }\textbf {\bibinfo {volume} {641}},\ \bibinfo {pages} {A6} (\bibinfo {year} {2020})},\ \bibinfo {note} {[Erratum: Astron.Astrophys. 652, C4 (2021)]},\ \Eprint {http://arxiv.org/abs/1807.06209} {arXiv:1807.06209 [astro-ph.CO]} \BibitemShut {NoStop}%
\bibitem [{\citenamefont {Rice}\ and\ \citenamefont {Zhang}(2017)}]{rice_cosmological_2017}%
  \BibitemOpen
  \bibfield  {author} {\bibinfo {author} {\bibfnamefont {Jared~R.}\ \bibnamefont {Rice}}\ and\ \bibinfo {author} {\bibfnamefont {Bing}\ \bibnamefont {Zhang}},\ }\bibfield  {title} {\enquote {\bibinfo {title} {{Cosmological evolution of primordial black holes}},}\ }\href {\doibase 10.1016/j.jheap.2017.02.002} {\bibfield  {journal} {\bibinfo  {journal} {JHEAp}\ }\textbf {\bibinfo {volume} {13-14}},\ \bibinfo {pages} {22--31} (\bibinfo {year} {2017})},\ \Eprint {http://arxiv.org/abs/1702.08069} {arXiv:1702.08069 [astro-ph.HE]} \BibitemShut {NoStop}%
\bibitem [{\citenamefont {De~Luca}\ \emph {et~al.}(2020{\natexlab{b}})\citenamefont {De~Luca}, \citenamefont {Franciolini}, \citenamefont {Pani},\ and\ \citenamefont {Riotto}}]{de_luca_constraints_2020}%
  \BibitemOpen
  \bibfield  {author} {\bibinfo {author} {\bibfnamefont {V.}~\bibnamefont {De~Luca}}, \bibinfo {author} {\bibfnamefont {G.}~\bibnamefont {Franciolini}}, \bibinfo {author} {\bibfnamefont {P.}~\bibnamefont {Pani}}, \ and\ \bibinfo {author} {\bibfnamefont {A.}~\bibnamefont {Riotto}},\ }\bibfield  {title} {\enquote {\bibinfo {title} {{Constraints on Primordial Black Holes: the Importance of Accretion}},}\ }\href {\doibase 10.1103/PhysRevD.102.043505} {\bibfield  {journal} {\bibinfo  {journal} {Phys. Rev. D}\ }\textbf {\bibinfo {volume} {102}},\ \bibinfo {pages} {043505} (\bibinfo {year} {2020}{\natexlab{b}})},\ \Eprint {http://arxiv.org/abs/2003.12589} {arXiv:2003.12589 [astro-ph.CO]} \BibitemShut {NoStop}%
\bibitem [{\citenamefont {De~Luca}\ and\ \citenamefont {Bellomo}(2023)}]{de_luca_accretion_2024}%
  \BibitemOpen
  \bibfield  {author} {\bibinfo {author} {\bibfnamefont {Valerio}\ \bibnamefont {De~Luca}}\ and\ \bibinfo {author} {\bibfnamefont {Nicola}\ \bibnamefont {Bellomo}},\ }\bibfield  {title} {\enquote {\bibinfo {title} {{The accretion, emission, mass and spin evolution of primordial black holes}},}\ }\href@noop {} {\  (\bibinfo {year} {2023})},\ \Eprint {http://arxiv.org/abs/2312.14097} {arXiv:2312.14097 [astro-ph.CO]} \BibitemShut {NoStop}%
\bibitem [{Bal()}]{BallisticNote}%
  \BibitemOpen
  \href@noop {} {}\bibinfo {note} {The neutrinos propagate radially outward with ballistic motion because they are uncharged---and therefore are not influenced by galactic magnetic fields---and the cross-section for interaction with the interstellar medium is negligible. The thickness of the ``shell'' of propagating neutrinos is set by the PBH lifetime (see column 4 of Table~\ref{table:PBHInfo}), and in the cases of interest, $\tau_0 \lesssim {\cal O} (10^{-5} \, {\rm s})$. We therefore can assume that all emitted neutrinos from one explosion would strike a detector at the same time.}\BibitemShut {Stop}%
\bibitem [{\citenamefont {Naab}\ \emph {et~al.}(2023)\citenamefont {Naab}, \citenamefont {Ganster},\ and\ \citenamefont {Zhang}}]{naab_measurement_2023}%
  \BibitemOpen
  \bibfield  {author} {\bibinfo {author} {\bibfnamefont {Richard}\ \bibnamefont {Naab}}, \bibinfo {author} {\bibfnamefont {Erik}\ \bibnamefont {Ganster}}, \ and\ \bibinfo {author} {\bibfnamefont {Zelong}\ \bibnamefont {Zhang}} (\bibinfo {collaboration} {IceCube}),\ }\bibfield  {title} {\enquote {\bibinfo {title} {{Measurement of the astrophysical diffuse neutrino flux in a combined fit of IceCube's high energy neutrino data}},}\ }in\ \href@noop {} {\emph {\bibinfo {booktitle} {{38th International Cosmic Ray Conference}}}}\ (\bibinfo {year} {2023})\ \Eprint {http://arxiv.org/abs/2308.00191} {arXiv:2308.00191 [astro-ph.HE]} \BibitemShut {NoStop}%
\bibitem [{Tes()}]{TestNote}%
  \BibitemOpen
  \href@noop {} {}\bibinfo {note} {We calculate $\Phi_b$ and $\sigma_{\Phi_b}$ using standard error propagation, given the values for $E_{\rm break}$, $\gamma_2$, and $\Phi_{\nu}^{\rm IC} (100 \,{\rm TeV})$ reported in Ref.~\cite{naab_measurement_2023}. Next we note that when extrapolating the flux to energies $E_\nu > E_{\rm break}$, the relevant quantities obey $\sigma_{\Phi (E_\nu)} / \Phi_\nu (E_{\nu}) \geq \sigma_{\Phi_b} / \Phi_b$. Since the variance is dominated by the uncertainty in $E_{\rm break}$, we make the conservative estimate and adopt $\sigma_{\Phi (E_\nu)} / \Phi (E_{\nu}) = \sigma_{\Phi_b} / \Phi_b.$}\BibitemShut {NoStop}%
\bibitem [{\citenamefont {Navarro}\ \emph {et~al.}(1996)\citenamefont {Navarro}, \citenamefont {Frenk},\ and\ \citenamefont {White}}]{navarro_structure_1996}%
  \BibitemOpen
  \bibfield  {author} {\bibinfo {author} {\bibfnamefont {Julio~F.}\ \bibnamefont {Navarro}}, \bibinfo {author} {\bibfnamefont {Carlos~S.}\ \bibnamefont {Frenk}}, \ and\ \bibinfo {author} {\bibfnamefont {Simon D.~M.}\ \bibnamefont {White}},\ }\bibfield  {title} {\enquote {\bibinfo {title} {{The Structure of cold dark matter halos}},}\ }\href {\doibase 10.1086/177173} {\bibfield  {journal} {\bibinfo  {journal} {Astrophys.~J.}\ }\textbf {\bibinfo {volume} {462}},\ \bibinfo {pages} {563--575} (\bibinfo {year} {1996})},\ \Eprint {http://arxiv.org/abs/astro-ph/9508025} {arXiv:astro-ph/9508025} \BibitemShut {NoStop}%
\bibitem [{\citenamefont {{Binney}}\ and\ \citenamefont {{Wong}}(2017)}]{binney_modelling_2017}%
  \BibitemOpen
  \bibfield  {author} {\bibinfo {author} {\bibfnamefont {James}\ \bibnamefont {{Binney}}}\ and\ \bibinfo {author} {\bibfnamefont {Leong~Khim}\ \bibnamefont {{Wong}}},\ }\bibfield  {title} {\enquote {\bibinfo {title} {{Modelling the Milky Way's globular cluster system}},}\ }\href {\doibase 10.1093/mnras/stx234} {\bibfield  {journal} {\bibinfo  {journal} {Mon.~Not.~Roy.~Astron.~Soc.}\ }\textbf {\bibinfo {volume} {467}},\ \bibinfo {pages} {2446--2457} (\bibinfo {year} {2017})},\ \Eprint {http://arxiv.org/abs/1701.06995} {arXiv:1701.06995 [astro-ph.GA]} \BibitemShut {NoStop}%
\bibitem [{\citenamefont {Posti}\ and\ \citenamefont {Helmi}(2019)}]{posti_mass_2019}%
  \BibitemOpen
  \bibfield  {author} {\bibinfo {author} {\bibfnamefont {Lorenzo}\ \bibnamefont {Posti}}\ and\ \bibinfo {author} {\bibfnamefont {Amina}\ \bibnamefont {Helmi}},\ }\bibfield  {title} {\enquote {\bibinfo {title} {{Mass and shape of the Milky Way\textquoteright{}s dark matter halo with globular clusters from Gaia and Hubble}},}\ }\href {\doibase 10.1051/0004-6361/201833355} {\bibfield  {journal} {\bibinfo  {journal} {Astron. Astrophys.}\ }\textbf {\bibinfo {volume} {621}},\ \bibinfo {pages} {A56} (\bibinfo {year} {2019})},\ \Eprint {http://arxiv.org/abs/1805.01408} {arXiv:1805.01408 [astro-ph.GA]} \BibitemShut {NoStop}%
\bibitem [{\citenamefont {Cerde\~{n}o}\ and\ \citenamefont {Green}(2010)}]{cerdeno_particle_2010}%
  \BibitemOpen
  \bibfield  {author} {\bibinfo {author} {\bibfnamefont {D.~G.}\ \bibnamefont {Cerde\~{n}o}}\ and\ \bibinfo {author} {\bibfnamefont {A.~M.}\ \bibnamefont {Green}},\ }\enquote {\bibinfo {title} {Direct detection of wimps},}\ in\ \href {\doibase 10.1017/CBO9780511770739.018} {\emph {\bibinfo {booktitle} {Particle Dark Matter: Observations, Models and Searches}}},\ \bibinfo {editor} {edited by\ \bibinfo {editor} {\bibfnamefont {Gianfranco}\ \bibnamefont {Bertone}}}\ (\bibinfo  {publisher} {Cambridge University Press},\ \bibinfo {year} {2010})\ p.\ \bibinfo {pages} {347–369},\ \Eprint {http://arxiv.org/abs/1002.1912} {arXiv:1002.1912 [astro-ph.CO]} \BibitemShut {NoStop}%
\bibitem [{\citenamefont {Choi}\ \emph {et~al.}(2014)\citenamefont {Choi}, \citenamefont {Rott},\ and\ \citenamefont {Itow}}]{choi_impact_2014}%
  \BibitemOpen
  \bibfield  {author} {\bibinfo {author} {\bibfnamefont {K.}~\bibnamefont {Choi}}, \bibinfo {author} {\bibfnamefont {Carsten}\ \bibnamefont {Rott}}, \ and\ \bibinfo {author} {\bibfnamefont {Yoshitaka}\ \bibnamefont {Itow}},\ }\bibfield  {title} {\enquote {\bibinfo {title} {{Impact of the dark matter velocity distribution on capture rates in the Sun}},}\ }\href {\doibase 10.1088/1475-7516/2014/05/049} {\bibfield  {journal} {\bibinfo  {journal} {JCAP}\ }\textbf {\bibinfo {volume} {05}},\ \bibinfo {pages} {049} (\bibinfo {year} {2014})},\ \Eprint {http://arxiv.org/abs/1312.0273} {arXiv:1312.0273 [astro-ph.HE]} \BibitemShut {NoStop}%
\bibitem [{\citenamefont {Mosbech}\ and\ \citenamefont {Picker}(2022)}]{mosbech_effects_2022}%
  \BibitemOpen
  \bibfield  {author} {\bibinfo {author} {\bibfnamefont {Markus~R.}\ \bibnamefont {Mosbech}}\ and\ \bibinfo {author} {\bibfnamefont {Zachary S.~C.}\ \bibnamefont {Picker}},\ }\bibfield  {title} {\enquote {\bibinfo {title} {{Effects of Hawking evaporation on PBH distributions}},}\ }\href {\doibase 10.21468/SciPostPhys.13.4.100} {\bibfield  {journal} {\bibinfo  {journal} {SciPost Phys.}\ }\textbf {\bibinfo {volume} {13}},\ \bibinfo {pages} {100} (\bibinfo {year} {2022})},\ \Eprint {http://arxiv.org/abs/2203.05743} {arXiv:2203.05743 [astro-ph.HE]} \BibitemShut {NoStop}%
\bibitem [{\citenamefont {Cang}\ \emph {et~al.}(2022)\citenamefont {Cang}, \citenamefont {Gao},\ and\ \citenamefont {Ma}}]{Cang:2021owu}%
  \BibitemOpen
  \bibfield  {author} {\bibinfo {author} {\bibfnamefont {Junsong}\ \bibnamefont {Cang}}, \bibinfo {author} {\bibfnamefont {Yu}~\bibnamefont {Gao}}, \ and\ \bibinfo {author} {\bibfnamefont {Yin-Zhe}\ \bibnamefont {Ma}},\ }\bibfield  {title} {\enquote {\bibinfo {title} {{21-cm constraints on spinning primordial black holes}},}\ }\href {\doibase 10.1088/1475-7516/2022/03/012} {\bibfield  {journal} {\bibinfo  {journal} {JCAP}\ }\textbf {\bibinfo {volume} {03}},\ \bibinfo {pages} {012} (\bibinfo {year} {2022})},\ \Eprint {http://arxiv.org/abs/2108.13256} {arXiv:2108.13256 [astro-ph.CO]} \BibitemShut {NoStop}%
\bibitem [{Pag()}]{PageNote}%
  \BibitemOpen
  \href@noop {} {}\bibinfo {note} {The Page factor $f(M)$ is ${\cal{O}}(1)$ for $M\geq 10^{15} {\rm g}$; it reaches a maximum of $\sim 15$ for $M\leq 10^{11} {\rm g}$. $f(M)$ varies significantly with mass beginning around $M=10^{15} {\rm g}$. This means that for PBHs that form with $M_i \gtrsim M_*$, $f(M)$ is nearly constant throughout the present age of the universe $t_0$. For further discussion, see Ref.~\cite{Klipfel:2025bvh}.}\BibitemShut {Stop}%
\bibitem [{\citenamefont {Niemeyer}\ and\ \citenamefont {Jedamzik}(1998)}]{Niemeyer:1997mt}%
  \BibitemOpen
  \bibfield  {author} {\bibinfo {author} {\bibfnamefont {Jens~C.}\ \bibnamefont {Niemeyer}}\ and\ \bibinfo {author} {\bibfnamefont {K.}~\bibnamefont {Jedamzik}},\ }\bibfield  {title} {\enquote {\bibinfo {title} {{Near-critical gravitational collapse and the initial mass function of primordial black holes}},}\ }\href {\doibase 10.1103/PhysRevLett.80.5481} {\bibfield  {journal} {\bibinfo  {journal} {Phys. Rev. Lett.}\ }\textbf {\bibinfo {volume} {80}},\ \bibinfo {pages} {5481--5484} (\bibinfo {year} {1998})},\ \Eprint {http://arxiv.org/abs/astro-ph/9709072} {arXiv:astro-ph/9709072} \BibitemShut {NoStop}%
\bibitem [{\citenamefont {Green}\ and\ \citenamefont {Liddle}(1999)}]{Green:1999xm}%
  \BibitemOpen
  \bibfield  {author} {\bibinfo {author} {\bibfnamefont {Anne~M.}\ \bibnamefont {Green}}\ and\ \bibinfo {author} {\bibfnamefont {Andrew~R.}\ \bibnamefont {Liddle}},\ }\bibfield  {title} {\enquote {\bibinfo {title} {{Critical collapse and the primordial black hole initial mass function}},}\ }\href {\doibase 10.1103/PhysRevD.60.063509} {\bibfield  {journal} {\bibinfo  {journal} {Phys. Rev. D}\ }\textbf {\bibinfo {volume} {60}},\ \bibinfo {pages} {063509} (\bibinfo {year} {1999})},\ \Eprint {http://arxiv.org/abs/astro-ph/9901268} {arXiv:astro-ph/9901268} \BibitemShut {NoStop}%
\bibitem [{\citenamefont {K\"uhnel}\ \emph {et~al.}(2016)\citenamefont {K\"uhnel}, \citenamefont {Rampf},\ and\ \citenamefont {Sandstad}}]{Kuhnel:2015vtw}%
  \BibitemOpen
  \bibfield  {author} {\bibinfo {author} {\bibfnamefont {Florian}\ \bibnamefont {K\"uhnel}}, \bibinfo {author} {\bibfnamefont {Cornelius}\ \bibnamefont {Rampf}}, \ and\ \bibinfo {author} {\bibfnamefont {Marit}\ \bibnamefont {Sandstad}},\ }\bibfield  {title} {\enquote {\bibinfo {title} {{Effects of Critical Collapse on Primordial Black-Hole Mass Spectra}},}\ }\href {\doibase 10.1140/epjc/s10052-016-3945-8} {\bibfield  {journal} {\bibinfo  {journal} {Eur. Phys. J. C}\ }\textbf {\bibinfo {volume} {76}},\ \bibinfo {pages} {93} (\bibinfo {year} {2016})},\ \Eprint {http://arxiv.org/abs/1512.00488} {arXiv:1512.00488 [astro-ph.CO]} \BibitemShut {NoStop}%
\bibitem [{\citenamefont {Gow}\ \emph {et~al.}(2022)\citenamefont {Gow}, \citenamefont {Byrnes},\ and\ \citenamefont {Hall}}]{gow_accurate_2022}%
  \BibitemOpen
  \bibfield  {author} {\bibinfo {author} {\bibfnamefont {Andrew~D.}\ \bibnamefont {Gow}}, \bibinfo {author} {\bibfnamefont {Christian~T.}\ \bibnamefont {Byrnes}}, \ and\ \bibinfo {author} {\bibfnamefont {Alex}\ \bibnamefont {Hall}},\ }\bibfield  {title} {\enquote {\bibinfo {title} {{Accurate model for the primordial black hole mass distribution from a peak in the power spectrum}},}\ }\href {\doibase 10.1103/PhysRevD.105.023503} {\bibfield  {journal} {\bibinfo  {journal} {Phys. Rev. D}\ }\textbf {\bibinfo {volume} {105}},\ \bibinfo {pages} {023503} (\bibinfo {year} {2022})},\ \Eprint {http://arxiv.org/abs/2009.03204} {arXiv:2009.03204 [astro-ph.CO]} \BibitemShut {NoStop}%
\bibitem [{Phi()}]{PhiPsiNote}%
  \BibitemOpen
  \href@noop {} {}\bibinfo {note} {To relate $\phi_{\rm GCC} (M_i)$ to the mass distribution $\psi_{\rm GCC} (M_i)$ considered in Refs.~\cite{gow_accurate_2022,gorton_how_2024}, we use the fact that $n_{\rm PBH} \simeq \rho_{\rm PBH} / \bar{M}$ for these distributions and ensure that $\phi_{\rm GCC} (M_i)$ is properly normalized.}\BibitemShut {Stop}%
\bibitem [{\citenamefont {Vitagliano}\ \emph {et~al.}(2020)\citenamefont {Vitagliano}, \citenamefont {Tamborra},\ and\ \citenamefont {Raffelt}}]{Vitagliano:2019yzm}%
  \BibitemOpen
  \bibfield  {author} {\bibinfo {author} {\bibfnamefont {Edoardo}\ \bibnamefont {Vitagliano}}, \bibinfo {author} {\bibfnamefont {Irene}\ \bibnamefont {Tamborra}}, \ and\ \bibinfo {author} {\bibfnamefont {Georg}\ \bibnamefont {Raffelt}},\ }\bibfield  {title} {\enquote {\bibinfo {title} {{Grand Unified Neutrino Spectrum at Earth: Sources and Spectral Components}},}\ }\href {\doibase 10.1103/RevModPhys.92.045006} {\bibfield  {journal} {\bibinfo  {journal} {Rev. Mod. Phys.}\ }\textbf {\bibinfo {volume} {92}},\ \bibinfo {pages} {45006} (\bibinfo {year} {2020})},\ \Eprint {http://arxiv.org/abs/1910.11878} {arXiv:1910.11878 [astro-ph.HE]} \BibitemShut {NoStop}%
\bibitem [{\citenamefont {Ivanchik}\ \emph {et~al.}(2024)\citenamefont {Ivanchik}, \citenamefont {Kurichin},\ and\ \citenamefont {Yurchenko}}]{Ivanchik:2024mqq}%
  \BibitemOpen
  \bibfield  {author} {\bibinfo {author} {\bibfnamefont {Alexandre~V.}\ \bibnamefont {Ivanchik}}, \bibinfo {author} {\bibfnamefont {Oleg~A.}\ \bibnamefont {Kurichin}}, \ and\ \bibinfo {author} {\bibfnamefont {Vlad~Yu.}\ \bibnamefont {Yurchenko}},\ }\bibfield  {title} {\enquote {\bibinfo {title} {{Neutrino at Different Epochs of the Friedmann Universe}},}\ }\href {\doibase 10.3390/universe10040169} {\bibfield  {journal} {\bibinfo  {journal} {Universe}\ }\textbf {\bibinfo {volume} {10}},\ \bibinfo {pages} {169} (\bibinfo {year} {2024})},\ \Eprint {http://arxiv.org/abs/2404.07081} {arXiv:2404.07081 [astro-ph.CO]} \BibitemShut {NoStop}%
\bibitem [{\citenamefont {Yang}\ \emph {et~al.}(2024)\citenamefont {Yang}, \citenamefont {Wang}, \citenamefont {Zhao},\ and\ \citenamefont {Zhang}}]{Yang:2024vij}%
  \BibitemOpen
  \bibfield  {author} {\bibinfo {author} {\bibfnamefont {Chen}\ \bibnamefont {Yang}}, \bibinfo {author} {\bibfnamefont {Sai}\ \bibnamefont {Wang}}, \bibinfo {author} {\bibfnamefont {Meng-Lin}\ \bibnamefont {Zhao}}, \ and\ \bibinfo {author} {\bibfnamefont {Xin}\ \bibnamefont {Zhang}},\ }\bibfield  {title} {\enquote {\bibinfo {title} {{Search for the Hawking radiation of primordial black holes: prospective sensitivity of LHAASO}},}\ }\href {\doibase 10.1088/1475-7516/2024/10/083} {\bibfield  {journal} {\bibinfo  {journal} {JCAP}\ }\textbf {\bibinfo {volume} {10}},\ \bibinfo {pages} {083} (\bibinfo {year} {2024})},\ \Eprint {http://arxiv.org/abs/2408.10897} {arXiv:2408.10897 [astro-ph.HE]} \BibitemShut {NoStop}%
\bibitem [{\citenamefont {Qin}\ \emph {et~al.}(2023)\citenamefont {Qin}, \citenamefont {Geller}, \citenamefont {Balaji}, \citenamefont {McDonough},\ and\ \citenamefont {Kaiser}}]{Qin:2023lgo}%
  \BibitemOpen
  \bibfield  {author} {\bibinfo {author} {\bibfnamefont {Wenzer}\ \bibnamefont {Qin}}, \bibinfo {author} {\bibfnamefont {Sarah~R.}\ \bibnamefont {Geller}}, \bibinfo {author} {\bibfnamefont {Shyam}\ \bibnamefont {Balaji}}, \bibinfo {author} {\bibfnamefont {Evan}\ \bibnamefont {McDonough}}, \ and\ \bibinfo {author} {\bibfnamefont {David~I.}\ \bibnamefont {Kaiser}},\ }\bibfield  {title} {\enquote {\bibinfo {title} {{Planck constraints and gravitational wave forecasts for primordial black hole dark matter seeded by multifield inflation}},}\ }\href {\doibase 10.1103/PhysRevD.108.043508} {\bibfield  {journal} {\bibinfo  {journal} {Phys. Rev. D}\ }\textbf {\bibinfo {volume} {108}},\ \bibinfo {pages} {043508} (\bibinfo {year} {2023})},\ \Eprint {http://arxiv.org/abs/2303.02168} {arXiv:2303.02168 [astro-ph.CO]} \BibitemShut {NoStop}%
\bibitem [{\citenamefont {Tran}\ \emph {et~al.}(2024)\citenamefont {Tran}, \citenamefont {Geller}, \citenamefont {Lehmann},\ and\ \citenamefont {Kaiser}}]{Tran:2023jci}%
  \BibitemOpen
  \bibfield  {author} {\bibinfo {author} {\bibfnamefont {Tung~X.}\ \bibnamefont {Tran}}, \bibinfo {author} {\bibfnamefont {Sarah~R.}\ \bibnamefont {Geller}}, \bibinfo {author} {\bibfnamefont {Benjamin~V.}\ \bibnamefont {Lehmann}}, \ and\ \bibinfo {author} {\bibfnamefont {David~I.}\ \bibnamefont {Kaiser}},\ }\bibfield  {title} {\enquote {\bibinfo {title} {{Close encounters of the primordial kind: A new observable for primordial black holes as dark matter}},}\ }\href {\doibase 10.1103/PhysRevD.110.063533} {\bibfield  {journal} {\bibinfo  {journal} {Phys. Rev. D}\ }\textbf {\bibinfo {volume} {110}},\ \bibinfo {pages} {063533} (\bibinfo {year} {2024})},\ \Eprint {http://arxiv.org/abs/2312.17217} {arXiv:2312.17217 [astro-ph.CO]} \BibitemShut {NoStop}%
\bibitem [{\citenamefont {Cuadrat-Grzybowski}\ \emph {et~al.}(2024)\citenamefont {Cuadrat-Grzybowski}, \citenamefont {Clesse}, \citenamefont {Defraigne}, \citenamefont {Van~Camp},\ and\ \citenamefont {Bertrand}}]{Cuadrat-Grzybowski:2024uph}%
  \BibitemOpen
  \bibfield  {author} {\bibinfo {author} {\bibfnamefont {Michal}\ \bibnamefont {Cuadrat-Grzybowski}}, \bibinfo {author} {\bibfnamefont {S{\'e}bastien}\ \bibnamefont {Clesse}}, \bibinfo {author} {\bibfnamefont {Pascale}\ \bibnamefont {Defraigne}}, \bibinfo {author} {\bibfnamefont {Michel}\ \bibnamefont {Van~Camp}}, \ and\ \bibinfo {author} {\bibfnamefont {Bruno}\ \bibnamefont {Bertrand}},\ }\bibfield  {title} {\enquote {\bibinfo {title} {{Probing primordial black holes and dark matter clumps in the Solar System with gravimeter and Global Navigation Satellite Systems networks}},}\ }\href {\doibase 10.1103/PhysRevD.110.063029} {\bibfield  {journal} {\bibinfo  {journal} {Phys. Rev. D}\ }\textbf {\bibinfo {volume} {110}},\ \bibinfo {pages} {063029} (\bibinfo {year} {2024})},\ \Eprint {http://arxiv.org/abs/2403.14397} {arXiv:2403.14397 [astro-ph.CO]} \BibitemShut {NoStop}%
\bibitem [{\citenamefont {De~Lorenci}\ \emph {et~al.}(2025)\citenamefont {De~Lorenci}, \citenamefont {Kaiser}, \citenamefont {Peter}, \citenamefont {Ruiz},\ and\ \citenamefont {Wolfe}}]{DeLorenci:2025wbn}%
  \BibitemOpen
  \bibfield  {author} {\bibinfo {author} {\bibfnamefont {Vitorio~A.}\ \bibnamefont {De~Lorenci}}, \bibinfo {author} {\bibfnamefont {David~I.}\ \bibnamefont {Kaiser}}, \bibinfo {author} {\bibfnamefont {Patrick}\ \bibnamefont {Peter}}, \bibinfo {author} {\bibfnamefont {Lucas~S.}\ \bibnamefont {Ruiz}}, \ and\ \bibinfo {author} {\bibfnamefont {Noah~E.}\ \bibnamefont {Wolfe}},\ }\bibfield  {title} {\enquote {\bibinfo {title} {{Gravitational wave signals from primordial black holes orbiting solar-type stars}},}\ }\href@noop {} {\  (\bibinfo {year} {2025})},\ \Eprint {http://arxiv.org/abs/2504.07517} {arXiv:2504.07517 [gr-qc]} \BibitemShut {NoStop}%
\bibitem [{\citenamefont {Klipfel}\ \emph {et~al.}(2025)\citenamefont {Klipfel}, \citenamefont {Fisher},\ and\ \citenamefont {Kaiser}}]{Klipfel:2025bvh}%
  \BibitemOpen
  \bibfield  {author} {\bibinfo {author} {\bibfnamefont {Alexandra~P.}\ \bibnamefont {Klipfel}}, \bibinfo {author} {\bibfnamefont {Peter}\ \bibnamefont {Fisher}}, \ and\ \bibinfo {author} {\bibfnamefont {David~I.}\ \bibnamefont {Kaiser}},\ }\bibfield  {title} {\enquote {\bibinfo {title} {{Hawking Radiation Signatures from Primordial Black Holes Transiting the Inner Solar System: Prospects for Detection}},}\ }\href@noop {} {\  (\bibinfo {year} {2025})},\ \Eprint {http://arxiv.org/abs/2506.14041} {arXiv:2506.14041 [astro-ph.CO]} \BibitemShut {NoStop}%
\bibitem [{\citenamefont {Gibbons}(1975)}]{gibbons_vacuum_1975}%
  \BibitemOpen
  \bibfield  {author} {\bibinfo {author} {\bibfnamefont {G.~W.}\ \bibnamefont {Gibbons}},\ }\bibfield  {title} {\enquote {\bibinfo {title} {{Vacuum Polarization and the Spontaneous Loss of Charge by Black Holes}},}\ }\href {\doibase 10.1007/BF01609829} {\bibfield  {journal} {\bibinfo  {journal} {Commun. Math. Phys.}\ }\textbf {\bibinfo {volume} {44}},\ \bibinfo {pages} {245--264} (\bibinfo {year} {1975})}\BibitemShut {NoStop}%
\bibitem [{\citenamefont {Carter}(1974)}]{carter_charge_1974}%
  \BibitemOpen
  \bibfield  {author} {\bibinfo {author} {\bibfnamefont {B.}~\bibnamefont {Carter}},\ }\bibfield  {title} {\enquote {\bibinfo {title} {{Charge and particle conservation in black hole decay}},}\ }\href {\doibase 10.1103/PhysRevLett.33.558} {\bibfield  {journal} {\bibinfo  {journal} {Phys. Rev. Lett.}\ }\textbf {\bibinfo {volume} {33}},\ \bibinfo {pages} {558--561} (\bibinfo {year} {1974})}\BibitemShut {NoStop}%
\bibitem [{\citenamefont {Wu}\ and\ \citenamefont {Xu}(2025)}]{Wu:2024uxa}%
  \BibitemOpen
  \bibfield  {author} {\bibinfo {author} {\bibfnamefont {Quan-feng}\ \bibnamefont {Wu}}\ and\ \bibinfo {author} {\bibfnamefont {Xun-Jie}\ \bibnamefont {Xu}},\ }\bibfield  {title} {\enquote {\bibinfo {title} {{High-energy and ultra-high-energy neutrinos from Primordial Black Holes}},}\ }\href {\doibase 10.1088/1475-7516/2025/02/059} {\bibfield  {journal} {\bibinfo  {journal} {JCAP}\ }\textbf {\bibinfo {volume} {02}},\ \bibinfo {pages} {059} (\bibinfo {year} {2025})},\ \Eprint {http://arxiv.org/abs/2409.09468} {arXiv:2409.09468 [hep-ph]} \BibitemShut {NoStop}%
\bibitem [{\citenamefont {Khlopov}(2024)}]{Khlopov:2024nqp}%
  \BibitemOpen
  \bibfield  {author} {\bibinfo {author} {\bibfnamefont {Maxim}\ \bibnamefont {Khlopov}},\ }\bibfield  {title} {\enquote {\bibinfo {title} {{Primordial Black Hole Messenger of Dark Universe}},}\ }\href {\doibase 10.3390/sym16111487} {\bibfield  {journal} {\bibinfo  {journal} {Symmetry}\ }\textbf {\bibinfo {volume} {16}},\ \bibinfo {pages} {1487} (\bibinfo {year} {2024})}\BibitemShut {NoStop}%
\bibitem [{\citenamefont {Choi}\ \emph {et~al.}(2025)\citenamefont {Choi}, \citenamefont {Lkhagvadorj},\ and\ \citenamefont {Mahapatra}}]{Choi:2025hqt}%
  \BibitemOpen
  \bibfield  {author} {\bibinfo {author} {\bibfnamefont {Ki-Young}\ \bibnamefont {Choi}}, \bibinfo {author} {\bibfnamefont {Erdenebulgan}\ \bibnamefont {Lkhagvadorj}}, \ and\ \bibinfo {author} {\bibfnamefont {Satyabrata}\ \bibnamefont {Mahapatra}},\ }\bibfield  {title} {\enquote {\bibinfo {title} {{Cosmological Origin of the KM3-230213A event and associated Gravitational Waves}},}\ }\href@noop {} {\  (\bibinfo {year} {2025})},\ \Eprint {http://arxiv.org/abs/2503.22465} {arXiv:2503.22465 [hep-ph]} \BibitemShut {NoStop}%
\bibitem [{\citenamefont {Baker}\ \emph {et~al.}(2025)\citenamefont {Baker}, \citenamefont {Iguaz~Juan}, \citenamefont {Symons},\ and\ \citenamefont {Thamm}}]{Baker:2025cff}%
  \BibitemOpen
  \bibfield  {author} {\bibinfo {author} {\bibfnamefont {Michael~J.}\ \bibnamefont {Baker}}, \bibinfo {author} {\bibfnamefont {Joaquim}\ \bibnamefont {Iguaz~Juan}}, \bibinfo {author} {\bibfnamefont {Aidan}\ \bibnamefont {Symons}}, \ and\ \bibinfo {author} {\bibfnamefont {Andrea}\ \bibnamefont {Thamm}},\ }\bibfield  {title} {\enquote {\bibinfo {title} {{Explaining the PeV Neutrino Fluxes at KM3NeT and IceCube with Quasi-Extremal Primordial Black Holes}},}\ }\href@noop {} {\  (\bibinfo {year} {2025})},\ \Eprint {http://arxiv.org/abs/2505.22722} {arXiv:2505.22722 [hep-ph]} \BibitemShut {NoStop}%
\bibitem [{\citenamefont {Olinto}\ \emph {et~al.}(2025)\citenamefont {Olinto} \emph {et~al.}}]{Olinto:2025elp}%
  \BibitemOpen
  \bibfield  {author} {\bibinfo {author} {\bibfnamefont {Angela~V.}\ \bibnamefont {Olinto}} \emph {et~al.},\ }\bibfield  {title} {\enquote {\bibinfo {title} {{Prospects for PBR detection of KM3-230213A-like events}},}\ }in\ \href@noop {} {\emph {\bibinfo {booktitle} {{39th International Cosmic Ray Conference}}}}\ (\bibinfo {year} {2025})\ \Eprint {http://arxiv.org/abs/2507.02246} {arXiv:2507.02246 [astro-ph.HE]} \BibitemShut {NoStop}%
\end{thebibliography}

%

\clearpage

\setcounter{equation}{0}
\setcounter{table}{0}
\setcounter{figure}{0}
\renewcommand{\theequation}{S\arabic{equation}}
\renewcommand{\thetable}{SM\Roman{table}}

\onecolumngrid

\centerline{\large\bf Supplemental Materials:} 
\centerline{\large\bf Ultra-High-Energy Neutrinos from Primordial Black Holes}
\vskip 30pt





\twocolumngrid

\begin{table*}[h!]
\begin{ruledtabular}
\begin{tabular}{cccccc}
 & $E_{\rm dep} \, [ {\rm PeV}]$ & Date & $\Delta T \, [ {\rm yr}]$ & Ref. & $N_\nu$ \\
 \hline
 IC & $1.01 \pm 0.16$ & Aug 8 2011 & $1.69$ & \cite{icecube_collaboration_first_2013} & $1.40_{-0.37}^{+0.53} \times 10^{24}$ \\
 IC & $1.14 \pm 0.17$ & Jan 3 2012 & $1.69$ & \cite{icecube_collaboration_first_2013} & $1.14_{-0.29}^{+0.43} \times 10^{24}$ \\
 IC & $2.00 \pm 0.26$ & Dec 4 2012 & $2.71$ & \cite{icecube_collaboration_observation_2014} &$3.43_{-0.48}^{+0.95} \times 10^{23}$ \\
 IC & $4.5 \pm 1.2$ & Jun 11 2014 & $2.90$ & \cite{aartsen_observation_2016} & $7.12_{-2.58}^{+6.07} \times 10^{22}$ \\
 IC & $6.05 \pm 0.72$ & Dec 8 2016 & $4.6$ & \cite{icecube_collaboration_detection_2021} & $3.00_{-1.00}^{+1.45} \times 10^{22}$ \\
 KM3 & $120^{+ 110}_{-60}$ & Feb 13 2023 & $0.79$ & \cite{aiello_observation_2025} & $1.05_{-0.79}^{+2.96} \times 10^{20}$ \\
\end{tabular}
\end{ruledtabular}
\caption{\justifying All cosmogenic high-energy neutrino events with $Q \geq 1 \, {\rm PeV}$ reported by the IceCube Collaboration (IC) and the KM3NeT Collaboration (KM3). The reported confidence intervals are $68\%$. $E_{\rm dep}$ is the energy deposited in the detector and $\Delta T$ is the livetime---the duration over which the dataset that includes the neutrino event was collected. We compute $N_\nu$ from Eq.~(6) using $Q_{\rm min}$ as the lower bound on the reported confidence interval for $E_{\rm dep}$. }
\label{table:NuEvents}
\end{table*}


\begin{table*}[h!]
\caption{\label{table:DOF} \justifying Parameters for Standard Model particles necessary to compute primary Hawking emission rates and lifetimes. Particles are listed in order of increasing PBH temperature $T_H$. The effective mass $m_{{\rm eff}, j}$ is the mass (or energy scale) used for a particle when computing the Page factor. The Page coefficients are computed using primary emission power integrals from Ref.~\cite{macgibbon_quark-_1990}. We use constituent masses for up and down quarks and the QCD IR cutoff scale for gluons as recommended by Ref.~\cite{macgibbon_quark-_1991}. Column 4 lists the number of degrees of freedom, $g_j$, for each particle. Columns 5-7 report the PBH temperature $T_H$ at which the primary emission spectrum is peaked at the particle rest mass $m_{\text{eff}, j}$, the PBH mass $M_j$ that corresponds to temperature $T_H$, and the lifetime $\tau$ of a PBH with mass $M$ computed via $  \tau(M_i) \simeq 1.98\times10^{-34}M_i^3f(M_i)^{-1} \text{ yr}$ \cite{macgibbon_quark-_1991}. Column 8 reports the calculated coefficients for each particle used to compute the Page factor $f(M)$ in Eq.~(3).\\}
\begin{ruledtabular}
\begin{tabular}{lccccccc}
 Particle & Symbol & $m_{\text{eff}, j}$ (GeV) & $g_j$ & $T_H$ (GeV) & $M_j$ (g) & $\tau$ (yr) & Page coefficient $\mathcal{P}_j$\\
 \hline
photon & $\gamma$ & 0 & 2 & - & - & - & 0.120 \\
neutrino & $\nu, \bar{\nu}$ & 0 & 6 & - & - & - & 0.882 \\
electron & $e^{\pm}$ & $5.11\times10^{-4}$ & 4 & $1.13\times10^{-3}$ & $9.4\times10^{16}$ & $1.0\times10^{17}$  & 0.568\\
muon & $\mu^{\pm}$ & 0.106 & 4 & $0.0233$ & $4.5\times10^{14}$ & $9.7\times10^9$ & 0.568\\
up quark & $u, \bar{u}$ & 0.201  & 12 &  $0.0444$ & $2.4\times10^{14}$  & $1.3\times10^9$ & 1.70  \\
down quark & $d, \bar{d}$ &  0.479 & 12 & $0.104$  & $1.0\times10^{14}$  & $4.4\times10^7$  & 1.70  \\
gluon & $g$ & 0.65  & 16  & $0.108$  & $9.8\times10^{13}$  & $4.1\times10^7$  & 0.96  \\
strange quark & $s, \bar{s}$ &  0.96 & 12  & $0.212$  & $4.5\times10^{13}$  & $2.8\times10^6$ & 1.70  \\
charm quark & $c, \bar{c}$ & 1.28  & 12  & $0.283$  & $3.7\times10^{13}$  & $1.4\times10^6$  & 1.70 \\
tau & $\tau^{\pm}$ &  1.78 & 4  & $0.392$  & $2.6\times10^{13}$  & $4.4\times10^5$  & 0.568  \\
bottom quark & $b, \bar{b}$ & 4.18  & 12  & $0.922$  & $1.1\times10^{13}$  &  $2.5\times10^4$ & 1.70 \\
W boson & $W^{\pm}$ & 80.4  & 6  & $13.3$  & $7.9\times10^{11}$  & 7.9  & 0.36 \\
Z boson & $Z_0$ & 90.2  & 3  & $15.1$  & $7.0\times10^{11}$  & 5.5  & 0.18  \\
top quark & $t, \bar{t}$ & 173.1  &  12 &  $38.2$ & $2.8\times10^{11}$  & 0.33 & 1.70  \\
Higgs boson & $h$ & 124.1  &  1 & $46.6$  & $2.3\times10^{11}$  & 0.18 & 0.267  \\
 
\end{tabular}
\end{ruledtabular}
\end{table*}

\twocolumngrid

\section*{Expected Flux from Extragalactic PBH Explosions}

Here we consider contributions to the expected high-energy neutrino flux from PBH explosions that occur outside the Milky Way. As we will see, such contributions remain subdominant to the flux arising from explosions within the Galaxy.

We begin by estimating the expected present-day emission rate of neutrinos with $E_\nu \sim 1 \, {\rm PeV}$ from all explosions within the Milky Way:
\beq
\dot{N}_{\nu, \rm MW} = n_0 \, N_\nu (1 \, {\rm PeV}) \, V_{\rm eff} (r_{\rm vir}) ,
\label{dotNMWdef}
\eeq
where the effective volume is weighted by the modified NFW DM profile introduced in Eq.~(6):
\beq
V_{\rm eff} = 2 \pi \int_0^{r_{\rm vir}} r \, dr \int_{-r_{\rm vir}}^{r_{\rm vir}} dz \,\frac{ \rho_{\rm DM} (r, z)}{\rho_0} .
\label{Veffdef}
\eeq
Upon taking $N_\nu = 1.40 \times 10^{24}$ from Table~\ref{table:NuEvents} and using present-day explosion rate $n_0 = 1.41 \times 10^3 \,{\rm pc}^{-3} \, {\rm yr}^{-1}$, this yields $\dot{N}_{\nu, \rm MW} = 1.30 \times 10^{34} \, {\rm s}^{-1}$. Next we assume that all $\sim100$ billion galaxies within the observable Universe are distributed isotropically within a sphere whose radius is equal to the present radius of the observable Universe, $r_{\rm obs} = 14.3 \, {\rm Gpc}$, which corresponds to a number density of galaxies $n_{\rm gal} = 8.16\times10^{-21} \, {\rm pc}^{-3}$. If each galaxy emits the same rate of $Q \geq 1 \, {\rm PeV}$ neutrinos as the Milky Way, and (for simplicity) we neglect the effects of redshift on neutrino energies emitted at cosmological distances, then we can conservatively bound
\beq
\begin{split}
\Phi_\nu^{\rm xgal} (1 \, {\rm PeV}) &< \frac{ n_{\rm gal} \, \dot{N}_{\nu, \rm MW} \, r_{\rm obs} }{4\pi E_\nu} \\ 
&\simeq 5.96 \times 10^{-21} \, {\rm GeV}^{-1} \, {\rm cm}^{-2} \, {\rm s}^{-1} \, {\rm sr}^{-1}.
\end{split}
\label{PhixgalPeV}
\eeq
The total reported IceCube flux for the three neutrino flavors is \cite{naab_measurement_2023}
\beq
\begin{split}
&\Phi_\nu^{\rm IC} (1 \, {\rm PeV}) = 9.66_{-4.88}^{+5.49}\times10^{-21} \, {\rm GeV}^{-1} \, {\rm cm}^{-2} \, {\rm s}^{-1} \, {\rm sr}^{-1}\\
\end{split}
\label{PhiICPeV}
\eeq

Eqs.~(\ref{PhixgalPeV}), which is a conservative upper bound, and (\ref{PhiICPeV}) suggest that extragalactic PBH explosions should contribute a subdominant fraction of high-energy neutrino events detected on Earth with $E_{\rm dep} \geq 1 \, {\rm PeV}$. In fact, Eq.~(\ref{PhixgalPeV}) overestimates the expected flux from extragalactic PBH explosions that would contribute to measured events with $E_{\rm dep} \geq 1 \, {\rm PeV}$, since it neglects the redshifting of $E_\nu$ for neutrinos that originate from PBH explosions at cosmological distances. For a given measured energy $E_{\rm dep}$, the energy $E_\nu$ of an extragalactic neutrino emitted at redshift $z$ would need to satisfy $E_\nu \geq E_{\rm dep} (1 + z)$. As shown in column 6 of Table~\ref{table:NuEvents}, the total number of emitted neutrinos $N_\nu$ falls as $Q_{\rm peak} \sim E_\nu$ increases. We therefore conclude that extragalactic PBH explosions would contribute a subdominant proportion of detected neutrino events with $E_{\rm dep} \geq 1 \, {\rm PeV}$ in experiments like IceCube. This estimate further confirms that the inferred isotropic volumetric PBH explosion rate described in the main text, $n_0 = 1.41 \times 10^3 \, {\rm pc}^{-3} \, {\rm yr}^{-1}$, should remain compatible with the upper limit set by the HAWC Collaboration \cite{albert_constraining_2020}, even when considering contributions from extragalactic PBH explosions.

Next we consider the flux of ultra-high-energy neutrinos, like the recent KM3NeT event, that would be expected from extragalactic PBH explosions. We again assume an isotropic volumetric PBH explosion rate of $n_0 = 1.41 \times 10^3 \, {\rm pc}^{-3} \, {\rm yr}^{-1}$ and take $N_\nu = 4.02 \times 10^{20}$ from Table~\ref{table:NuEvents} for $E_{\rm dep} = 60 \, {\rm PeV}$. Proceeding as above, we may then bound
\beq
\Phi_\nu^{\rm xgal} (60 \, {\rm PeV}) < 2.85 \times 10^{-26} \, {\rm GeV}^{-1} \, {\rm cm}^{-2} \, {\rm s}^{-1} \, {\rm sr}^{-1}  ,
\label{Phixgal60PeV}
\eeq
upon making the simplifying assumption of neglecting the redshifting of $E_\nu$. We may compare the upper bound in Eq.~(\ref{Phixgal60PeV}) with the flux reported by the KM3NeT collaboration \cite{aiello_observation_2025},
\beq
\Phi_\nu^{\rm KM3} (60 \, {\rm PeV}) = 4.83 \times 10^{-23} \, {\rm GeV}^{-1} \, {\rm cm}^{-2} \, {\rm s}^{-1} \, {\rm sr}^{-1}.
\label{PhiKM360PeV}
\eeq
We therefore find $\Phi_\nu^{\rm KM3} (60 \, {\rm PeV}) \gg \Phi_\nu^{\rm xgal} (60 \, {\rm PeV})$, even for the conservative upper bound estimated in Eq.~(\ref{Phixgal60PeV}).

\end{document}